\documentclass[journal]{IEEEtran}
\usepackage{multirow}
\usepackage{graphicx}
\usepackage{amsmath}
\usepackage[psamsfonts]{amssymb}
\usepackage{amsxtra}
\usepackage{threeparttable}
\usepackage{amsfonts}
\usepackage{mathrsfs}
\usepackage{arydshln}
\usepackage{cite}
\usepackage{enumitem}
\usepackage{url}

\makeatletter
\let\MYcaption\@makecaption
\makeatother
\usepackage{subcaption}
\makeatletter
\let\@makecaption\MYcaption
\makeatother
\newcommand{\myvector}[1]{\boldsymbol{#1}}
\newcommand{\mymatrix}[1]{\boldsymbol{\mathrm{#1}}}
\newcommand{\myset}[1]{\mathrm{#1}}
\newcommand{\gaussian}[3]{\mathcal{N}(#1 | #2, #3)}
\newcommand{\betapdf}[3]{\mathcal{B}(#1 | #2, #3)}
\DeclareMathOperator*{\argmin}{arg\,min}
\DeclareMathOperator*{\argmax}{arg\,max}
\newcounter{num}
\newcommand{\rnum}[1]{\setcounter{num}{#1} \roman{num}}

\begin{document}
%
\title{Scene Segmentation-Based Luminance Adjustment\\for Multi-Exposure Image Fusion}
%
%
%

\author{Yuma~Kinoshita,~\IEEEmembership{Student~Member,~IEEE,}
        and~Hitoshi~Kiya,~\IEEEmembership{Fellow,~IEEE,}
\thanks{Y. Kinoshita and H. Kiya are with 
Tokyo Metropolitan University, Tokyo, Japan
e-mail: kinoshita-yuma@ed.tmu.ac.jp and kiya@tmu.ac.jp}
\thanks{Manuscript received April 19, 2005; revised August 26, 2015.}}

%
%

\markboth{Journal of \LaTeX\ Class Files,~Vol.~14, No.~8, August~2015}%
{Shell \MakeLowercase{\textit{et al.}}: Bare Demo of IEEEtran.cls for IEEE Journals}
%



\maketitle

\begin{abstract}
  We propose a novel method for adjusting luminance
  for multi-exposure image fusion.
  For the adjustment, two novel scene segmentation approaches based on
  luminance distribution are also proposed.
  Multi-exposure image fusion is a method for producing images
  that are expected to be more informative and perceptually appealing
  than any of the input ones,
  by directly fusing photos taken with different exposures.
  However, existing fusion methods often produce unclear fused images
  when input images do not have a sufficient number of different exposure levels.
  In this paper, we point out that adjusting the luminance of
  input images makes it possible
  to improve the quality of the final fused images.
  This insight is the basis of the proposed method.
  The proposed method enables us to produce high-quality images,
  even when undesirable inputs are given.
  Visual comparison results show that the proposed method can produce images
  that clearly represent a whole scene.
  In addition, multi-exposure image fusion with the proposed method outperforms
  state-of-the-art fusion methods in terms of MEF-SSIM,
  discrete entropy, tone mapped image quality
  index, and statistical naturalness.
\end{abstract}

\begin{IEEEkeywords}
  Multi-exposure image fusion, luminance adjustment,
  image enhancement, image segmentation
\end{IEEEkeywords}

%
\IEEEpeerreviewmaketitle

\section{Introduction}
  The low dynamic range (LDR) of the imaging sensors used in modern digital cameras is
  a major factor preventing cameras from capturing images as good as human vision.
  This is due to the limited dynamic range that imaging sensors have,
  and a single shutter speed that is utilized when we take photos.
  The limitations result in low-contrast images taken by digital cameras.
  To obtain images with better quality,
  numerous imaging and enhancement techniques based on a single image
  have been proposed
  \cite{zuiderveld1994contrast, wu2017contrast, kinoshita2017pseudo,
  su2017low, ren2018joint, chen2018learning}.
  However, these techniques cannot restore saturated pixel values
  in LDR images.

  Because of this situation, interest in multi-image-based imaging techniques
  \cite{hasinoff2016burst,goshtasby2005fusion,mertens2009exposure,saleem2012image,
  li2012fast,wang2015exposure,li2014selectively,sakai2015hybrid, nejati2017fast,
  prabhakar2017deepfuse, li2013image}
  has been increasing to overcome this constraint that
  single-image-based techniques have.
  Multi-exposure image fusion is a solution to the LDR problem of digital cameras.
  Fusion methods utilize a set of differently exposed images,
  called ``multi-exposure images,'' and fuse them to produce an image with high quality.
  Their development was inspired by high dynamic range (HDR) imaging techniques
  \cite{debevec1997recovering,reinhard2002photographic,oh2015robust,
  kinoshita2016remapping,kinoshita2017fast,kinoshita2017fast_trans,huo2016single,
  murofushi2013integer,murofushi2014integer,dobashi2014fixed}.
  The advantage of these methods, compared with HDR imaging techniques, is that
  they can eliminate three operations:
  generating HDR images, calibrating the camera response function (CRF),
  and preserving the exposure value of each photograph.
  In this paper, we focus on multi-exposure image fusion.

  However, to work well, existing fusion methods require
  two conditions \cite{kinoshita2018multi,kinoshita2018automatic}:
  input multi-exposure images include no moving artifacts such as blurring
  or misalignment among images,
  and input multi-exposure images clearly cover the dynamic range of a scene.
  Moreover, the conditions have a close relation each other.
  To capture multi-exposure images covering a high dynamic range,
  it is generally needed to shoot multiple times while changing the shutter speed.
  In this case, moving artifacts generally occur due to time lags of each shooting
  and long-shutter speeds to capture bright images.
  Therefore, the two conditions are not always satisfied at the same time
  in various practical situations.

  For eliminating moving artifacts among images,
  specialized camera systems \cite{nayar2000high, an2017single} and
  a number of robust fusion methods \cite{ma2017robust}
  have been studied.
  For specialized camera systems,
  spatially varying pixel exposure methods \cite{nayar2000high, an2017single}
  using a single sensor
  and new optical systems \cite{aggarwal2004split, tocch2011versatile}
  having a number of sensors have been proposed.
  These systems enable us to capture scenes at the same time
  by changing the amount of light absorbed at each pixel in the imaging sensor,
  or by varying the exposure for each sensor.
  However, these camera systems do not always capture images
  having a sufficient number of different exposure levels,
  due to the hardware limitation of camera systems.

  In contrast, for robust fusion methods against moving artifacts,
  input multi-exposure images taken with conventional cameras are first aligned,
  and then aligned images are fused.
  These methods effectively produce high-quality images
  even when scenes include moving objects,
  if well-exposed multi-exposure images are prepared.
  However, the quality of input multi-exposure images affects
  both the accuracy of image alignment and the performance of fusion.
  Hence, insufficient multi-exposure images,
  such as ones taken with fast-shutter-speed due to limitations of camera capability,
  make it difficult to produce high-quality results.
  Because of these situations,
  we aim to propose
  a scene segmentation-based luminance adjustment method,
  for improving the quality of insufficient multi-exposure images.

  In this paper, we first point out that
  it is possible to improve the quality of multi-exposure images
  by adjusting the luminance of images after photographing.
  Moreover, a scene segmentation-based luminance adjustment method is proposed
  on the basis of this insight.
  The proposed method enables us to produce high-quality images
  even when unclear input images are given.
  For the adjustment, two scene segmentation approaches are also proposed
  in order to automatically adjust input multi-exposure images
  so that they become suitable for multi-exposure image fusion.
  The former separates a scene in multi-exposure images,
  according to the luminance values of
  an input image having middle brightness.
  This approach has a closed-form,
  so it has a lower computational cost than typical segmentation methods.
  In the latter, which provides high-quality results,
  a scene is separated by considering
  the luminance distribution of all input images.
  The quality of fused images generally increases by using
  a large number of multi-exposure images,
  but the use of these images requires a large computational cost for fusion.
  In addition, the suitable number depends on a scene.
  For these reasons, the latter segmentation method automatically
  determines the suitable number of adjusted multi-exposure images,
  although most segmentation ones including $k$-means cannot.
  
  In simulations, we evaluate the effectiveness of the proposed method
  with various fusion methods
  in terms of visual and quantitative comparison.
  Visual comparison results show that the proposed method can produce images
  that clearly represent a whole scene.
  In addition, multi-exposure image fusion with the proposed method outperforms
  state-of-the-art fusion methods in terms of MEF-SSIM,
  discrete entropy, tone mapped image quality index, and statistical naturalness.
\section{Previous work}
  As mentioned,
  existing multi-exposure image fusion methods use differently exposed images,
  referred to as ``multi-exposure images.''
  Here, we summarize typical fusion methods and problems with them.

  Various research works on multi-exposure image fusion have so far been reported
  for capturing images of real scenes having a wide dynamic range
  without clipped blacks or whites
  \cite{goshtasby2005fusion,mertens2009exposure,saleem2012image,wang2015exposure,
  li2014selectively,sakai2015hybrid, nejati2017fast}.
  Many of the fusion methods provide the final fused image as a weighted average
  of input multi-exposure images.
  Mertens et al.\cite{mertens2009exposure} proposed a multi-scale fusion scheme in which
  contrast, color saturation, and well-exposedness measures are used
  for computing fusion weights.
  In the work by Nejati et al. \cite{nejati2017fast},
  base-detail decomposition is applied to each input image,
  and base and detail layers are then combined individually.
  An exposure fusion method based on sparse representation,
  which combines sparse coefficients with fused images,
  was also proposed in \cite{wang2015exposure}.
  Furthermore, a method that combines weighted average and
  sparse representation is presented in \cite{sakai2015hybrid}
  and is used
  to enhance image details.
  The existing methods work very well when the following two conditions are satisfied.
  \begin{itemize}[nosep]
    \item The scene is static, and the camera is tripod-mounted.
    \item Input multi-exposure images clearly cover the dynamic range of the scene.
  \end{itemize}
  Especially, photographing moving objects with an unstable camera is
  the most common scenario in imaging.
  Hence, moving artifacts such as motion blur and misalignment generally occur
  when the long exposure time is used to capture sufficient multi-exposure images.
  These moving artifacts result in ghost-like artifacts in the final fused images.
  
  Because of this,
  specialized camera systems and a lot of fusion methods
  have been proposed to eliminate ghost-like artifacts.
  Spatially varying pixel exposure methods \cite{nayar2000high, an2017single}
  and new optical systems \cite{aggarwal2004split, tocch2011versatile}
  have been studied to implement specialized camera systems.
  The former methods enable us to use conventional cameras
  for obtaining single-shot multi-exposure images
  by changing the amount of light absorbed at each pixel
  in the imaging sensor.
  The latter systems capture multi-exposure images with multiple sensors
  by splitting the light onto the sensors with different absorptive filters.
  However, the camera systems can capture a limited number of multi-exposure images
  due to the hardware limitation of the systems.
  For robust fusion methods against moving artifacts,
  Ma et al. \cite{ma2017robust} proposed
  a structural patch decomposition method for multi-exposure image fusion
  as one fusion method.
  By these research works, problems of the ghost-like artifacts
  are being solved,
  if well-exposed multi-exposure images are prepared.
  However, the quality of input multi-exposure images affects
  both the accuracy of image alignment and the performance of fusion.

\begin{figure}[!t]
  \centering
  \begin{subfigure}[t]{\hsize}
    \centering
    \includegraphics[width=0.30\columnwidth]{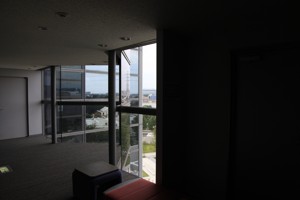}
    \includegraphics[width=0.30\columnwidth]{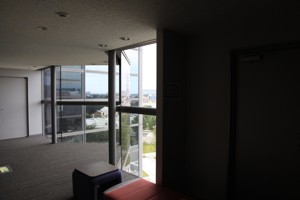}
    \includegraphics[width=0.30\columnwidth]{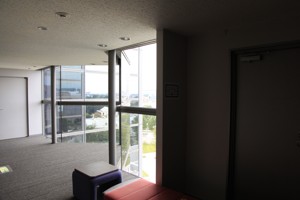}
    \caption{Input images ${\myvector{x}_n}$ ``Window''
      (exposed at $-1, 0$, and $+1 \mathrm{[EV]}$) \label{fig:window_input}}
  \end{subfigure}\\
  \begin{subfigure}[t]{\hsize}
    \centering
    \includegraphics[width=0.30\columnwidth]{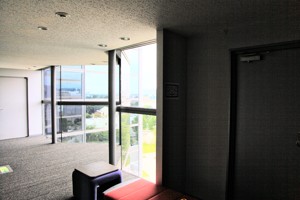}
    \includegraphics[width=0.30\columnwidth]{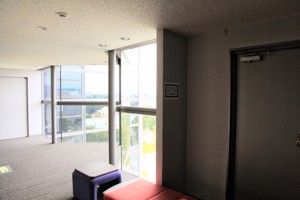}
    \includegraphics[width=0.30\columnwidth]{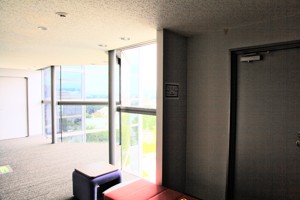}
    \caption{Example 1 of adjusted images\label{fig:window_enhanced_conv}}
  \end{subfigure}\\
  \begin{subfigure}[t]{\hsize}
    \centering
    \includegraphics[width=0.30\columnwidth]{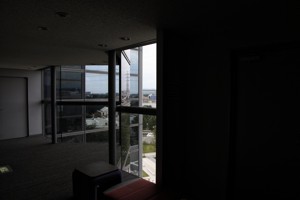}
    \includegraphics[width=0.30\columnwidth]{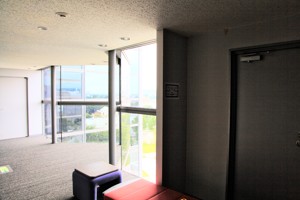}
    \includegraphics[width=0.30\columnwidth]{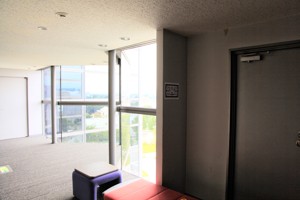}
    \caption{Example 2 of adjusted images \label{fig:window_enhanced_auto}}
  \end{subfigure}
  \caption{Examples of adjusted multi-exposure images.
    Luminance adjustment improves quality of multi-exposure images.}
  \label{fig:inputImages}
\end{figure}

\begin{figure}[!t]
  \centering
  \begin{subfigure}[t]{0.3\hsize}
    \centering
    \includegraphics[width=\columnwidth]{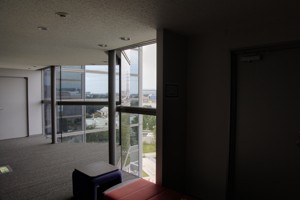}
    \caption{Input \label{fig:fused_orig}}
  \end{subfigure}
  \begin{subfigure}[t]{0.3\hsize}
    \centering
    \includegraphics[width=\columnwidth]{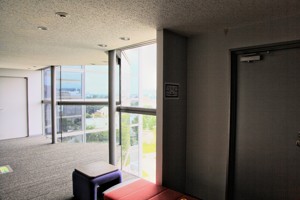}
    \caption{Fig. 1(b) \label{fig:fused_manual}}
  \end{subfigure}
  \begin{subfigure}[t]{0.3\hsize}
    \centering
    \includegraphics[width=\columnwidth]{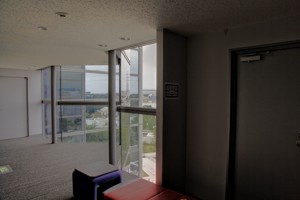}
    \caption{Fig. 1(c) \label{fig:fused_auto}}
  \end{subfigure}\\
  \caption{Examples of fused images.
    (a) Fused image from input multi-exposure images
    in Fig. {\protect\ref{fig:inputImages}(\subref{fig:window_input})},
    (b) fused image from adjusted multi-exposure images
    in Fig. {\protect\ref{fig:inputImages}(\subref{fig:window_enhanced_conv})},
    and
    (c) fused image from adjusted multi-exposure images
    in Fig. {\protect\ref{fig:inputImages}(\subref{fig:window_enhanced_auto})}.
    Quality of fused images is affected by degree of adjustment.
    \label{fig:fusedExamples}}
\end{figure}

  Here, we demonstrate that adjusting the luminance of unclear multi-exposure images
  affects the quality of the final fused images.
  Figures \ref{fig:inputImages} and \ref{fig:fusedExamples} show
  examples of adjusted multi-exposure images
  and fused images from these multi-exposure ones,
  respectively.
  These results indicate
  that adjusting the luminance makes it possible to improve the quality of fused images.
  In other words, even when appropriate exposure values are
  unknown at the time of photographing,
  the multi-exposure images can be improved by adjusting unclear input images.
  The quality of multi-exposure images depends on the degree of adjustment.

  Thus, we propose a novel luminance adjustment method,
  referred to as ``scene segmentation-based luminance adjustment'' (SSLA),
  for multi-exposure image fusion.

\section{Proposed scene-segmentation-based luminance adjustment}
  The proposed luminance adjustment method consists of three operations:
  local contrast enhancement, scene segmentation-based luminance scaling (SSLS),
  and tone mapping (see Fig. \ref{fig:proposedMEF}).
  An overview of the proposed method is first described,
  and SSLS is then explained.
\subsection{Notation}
  The following notations are used throughout this paper.
  \begin{itemize}[nosep]
    \item Lower case bold italic letters, e.g., $\myvector{x}$, denote
      vectors or vector-valued functions and are assumed to be column vectors.
    \item A superscript $\top$ denotes the transpose of a matrix or vector.
    \item The notation $(x_1, x_2, \cdots, x_N)$ denotes a $N$-dimensional row vector.
    \item The notation $\{x_1, x_2, \cdots, x_N\}$ denotes a set with $N$ elements.
      In situations where there is no ambiguity as to their elements,
      the simplified notation $\{x_n\}$ is used to denote the same set.
    \item The notation $p(x)$ denotes a probability density function of $x$.
    \item $U$ and $V$ are used to denote the width and height of input images, respectively.
    \item $\myset{P}$ denotes a set of all pixels, namely,
      $\myset{P} =
        \{(u, v)^\top | u \in \{1, 2, \cdots, U\} \land v \in \{1, 2, \cdots, V\}\}$.
    \item A pixel $\myvector{p}$ is written as $\myvector{p} = (u, v)^\top \in \myset{P}$.
    \item The number of input images is denoted by $N$.
    \item An input image is denoted by a vector-valued function
      $\myvector{x}: \myset{P} \to \mathbb{R}^3$,
      where its output means linear-RGB pixel values,
      namely, the RGB color space has a linear relationship with the CIE XYZ color space.
      Similarly, a fused image is denoted by $\myvector{y}: \myset{P} \to \mathbb{R}^3$.
      For example, linear-RGB pixel values are obtained
      by applying the inverse gamma correction to sRGB pixel values \cite{iec1999multimedia}
    \item The luminance of an image is denoted
      by the function $l: \myset{P} \to \mathbb{R}$,
      where its output corresponds to Y components of the CIE XYZ color space.
  \end{itemize}
\subsection{Overview}
\begin{figure*}[!t]
    \centering
    \includegraphics[clip, width=15cm]{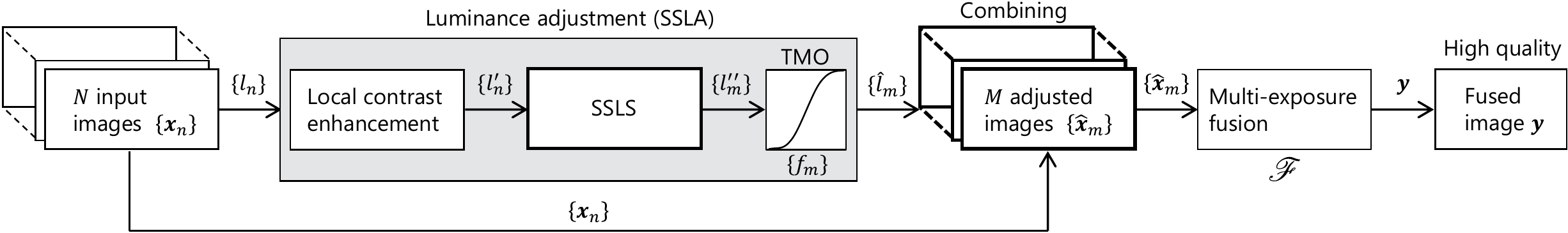}
    \caption{Use of proposed luminance adjustment method
      in multi-exposure image fusion.
      SSLS and way of producing adjusted images $\{\hat{\myvector{x}}_m\}$
      are our main contributions. \label{fig:proposedMEF}}
  \end{figure*}

  The use of the proposed method in multi-exposure image fusion is illustrated
  in Fig. \ref{fig:proposedMEF},
  where the main contributions of this paper are SSLS and the way of producing
  adjusted images $\{\hat{\myvector{x}}_m\}$.
  To enhance the quality of multi-exposure images, local contrast enhancement is applied to
  luminance $l_n$ calculated from the $n$-th input image $\myvector{x}_n$.
  Next, SSLS separates pixel set $\myset{P}$ into
  $M$-areas $\myset{P}_1, \myset{P}_2, \cdots, \myset{P}_M$
  according to luminance values,
  and luminance $l''_m$ for each area $\myset{P}_m$ is then obtained
  by scaling a set, $\{l'_n\}$, of the enhanced luminance.
  Here, the number $M$ of the scaled luminance,
  which equals the number of adjusted multi-exposure images,
  is generally different from the number $N$ of input images.
  In addition, tone mapping is applied to each scaled luminance $l''_m$
  in order to avoid the truncation of pixel values.
  The adjusted images $\{\hat{\myvector{x}}_m\}$ are generated
  by combining the mapped luminance $\{\hat{l}_m\}$
  and the input images $\{\myvector{x}_n\}$, and
  the resulting image $\myvector{y}$ is then fused from
  these adjusted ones $\{\hat{\myvector{x}}_m\}$,
  by using a multi-exposure image fusion method $\mathscr{F}$ such as weighted average.

  In Fig. \ref{fig:proposedMEF},
  the proposed luminance adjustment should be applied before fusion.
  If the luminance adjustment is applied after fusion,
  an additional fusion process is needed because
  the luminance adjustment generates multi-exposure images even when
  a single image is given as an input.
\subsubsection{Local contrast enhancement}
  If the input images do not represent a scene clearly,
  their contrast is lower than that of ideally exposed images.
  Therefore, the dodging and burning algorithm is used to enhance
  the local contrast\cite{huo2013dodging}.
  The luminance $l'_n$ enhanced by the algorithm is given by
  \begin{equation}
    l'_n (\myvector{p}) = \frac{l_n (\myvector{p})^2}{\bar{l}_n (\myvector{p})},\\
    \label{eq:dodgingAndBurning}
  \end{equation}
  where $\bar{l}_n (\myvector{p})$ is the local average of luminance $l_n (\myvector{p})$
  around pixel $\myvector{p} = (u, v)^\top$.
  It is obtained by applying a low-pass filter to $l_n (\myvector{p})$.
  Here, a bilateral filter is used for this purpose.

  $\bar{l}_n (\myvector{p})$ is calculated by using the bilateral filter:
  \begin{equation}
    \bar{l}_n (\myvector{p}) = \frac{
              \sum_{\myvector{p}' \in \myset{P}}
              l_n (\myvector{p}') g_{\sigma_1}(\| \myvector{p}' - \myvector{p} \|)
              g_{\sigma_2}(l_n (\myvector{p}') - l_n (\myvector{p}))}
              {\sum_{\myvector{p}' \in \myset{P}}
              g_{\sigma_1}(\| \myvector{p}' - \myvector{p} \|)
              g_{\sigma_2}(l_n (\myvector{p}') - l_n (\myvector{p}))},
    \label{eq:bilateral}
  \end{equation}
  where $g_\sigma(t)$ is a Gaussian function given by
  \begin{equation}
    g_\sigma(t) = \frac{1}{2 \pi \sigma^2} \exp \left( -\frac{t^2}{2\sigma^2} \right)
    \: \mathrm{for} \: t \in \mathbb{R}.
    \label{eq:gaussian}
  \end{equation}
  Parameters $\sigma_1 = 16$ and $\sigma_2 = 3/255$ are set
  in accordance with \cite{huo2013dodging},
  and a real-time implementation of the bilateral filter \cite{chen2007real}
  is utilized.
  
\subsubsection{\label{sec:tone_mapping} Tone mapping}
  Because the scaled luminance value $l''_m (\myvector{p})$ often exceeds
  the maximum value of the common image formats,
  pixel values might be lost due to truncation of the values.
  This problem is overcome by using a tone mapping operation
  to fit the adjusted luminance value into the interval $[0, 1]$.

  The luminance $\hat{l}_m (\myvector{p})$ of an adjusted multi-exposure image is obtained
  by applying a tone mapping operator $f_m$ to $l''_m (\myvector{p})$:
  \begin{equation}
    \hat{l}_m (\myvector{p}) = f_m(l''_m (\myvector{p})).
    \label{eq:TM}
  \end{equation}
  Reinhard's global operator is used here as the tone mapping operator $f_m$
  \cite{reinhard2002photographic}.
  
  Reinhard's global operator is given by
  \begin{equation}
    f_m(t) = \frac{t}{1 + t}\left(1 + \frac{t}{L^2_m} \right)
    \: \mathrm{for} \: t \in [0, \infty),
    \label{eq:reinhardTMO}
  \end{equation}
  where parameter $L_m > 0$ determines value $t$ to be $f_m(t) = 1$.
  Note that Reinhard's global operator $f_m$ is a monotonically increasing function.
  Here, setting $L_m$ as $L_m = \max l''_m (\myvector{p})$ for each $m$,
  we obtain $\hat{l}_m (\myvector{p}) \le 1$ for all $\myvector{p}$.
  Therefore, truncation of the luminance values can be prevented.
  In contrast, if it is preferable not to change luminance by tone mapping,
  setting $L_m$ as $L_m = 1$ is chosen.
  As described above, each of these two parameter settings has advantages and disadvantages.
  In this paper, parameter $L_m = \max l''_m (\myvector{p})$ is utilized.
\subsubsection{Fusion of enhanced multi-exposure images}
  Adjusted multi-exposure images ${\hat{\myvector{x}}_m}$ can be used as input for
  any existing multi-exposure image fusion methods \cite{mertens2009exposure,nejati2017fast}.
  A final image $\myvector{y}$ is produced as
  \begin{equation}
    \myvector{y} = \mathscr{F}(
      \hat{\myvector{x}}_1, \hat{\myvector{x}}_2, \cdots, \hat{\myvector{x}}_M),
    \label{eq:fusion}
  \end{equation}
  where $\mathscr{F}(\myvector{x}_1, \myvector{x}_2, \cdots, \myvector{x}_M)$
  indicates a function for fusing $M$ images
  $\myvector{x}_1, \myvector{x}_2, \cdots, \myvector{x}_M$
  into a single image.
\subsection{Scene segmentation-based luminance scaling}
\begin{figure}[!t]
    \centering
    \begin{subfigure}[t]{\hsize}
      \centering
      \includegraphics[clip, width=\columnwidth]{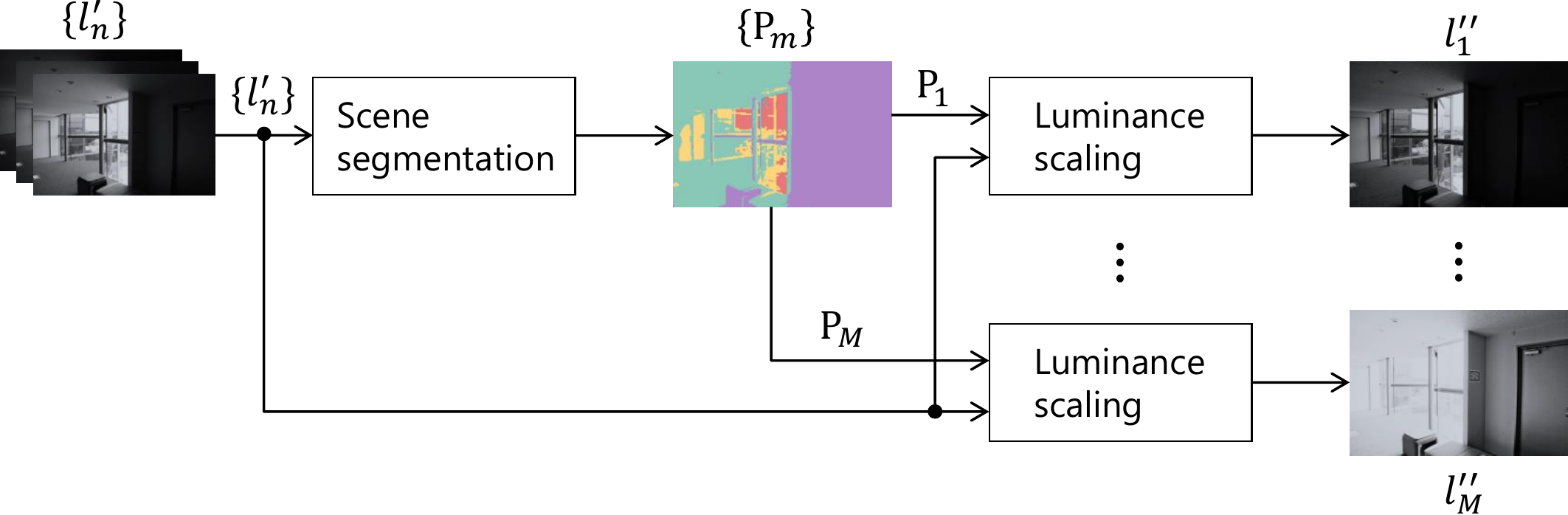}
      \caption{Flow of SSLS \label{fig:ssls_flow}}
    \end{subfigure}\\
    \begin{subfigure}[t]{\hsize}
      \centering
      \includegraphics[clip, width=\columnwidth]{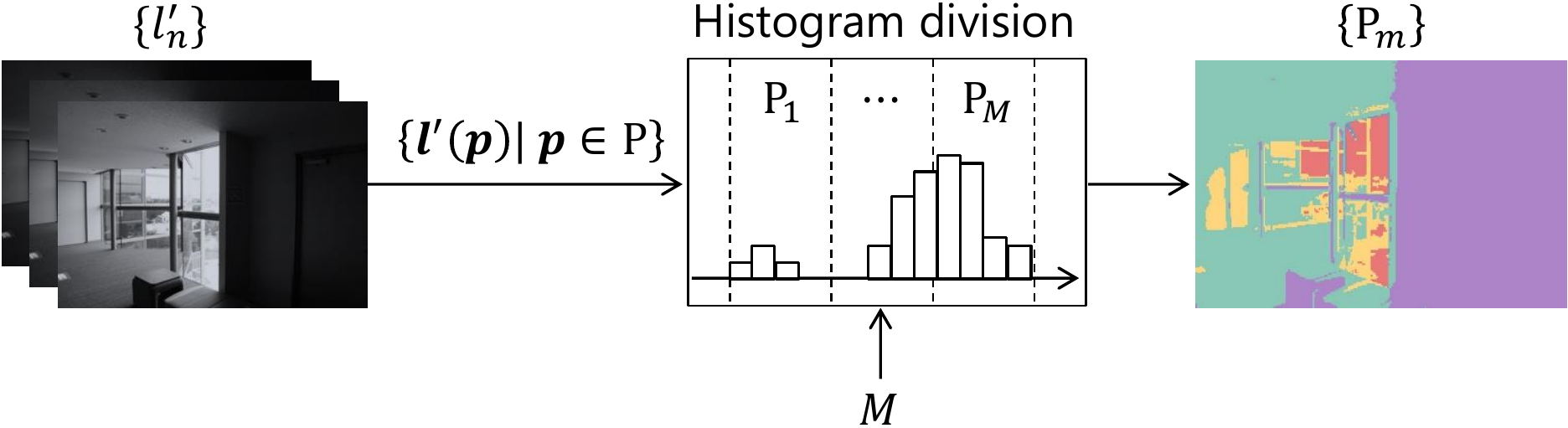}
      \caption{Scene segmentation (Approach 1) \label{fig:ssls_approach1}}
    \end{subfigure}\\
    \begin{subfigure}[t]{\hsize}
      \centering
      \includegraphics[clip, width=\columnwidth]{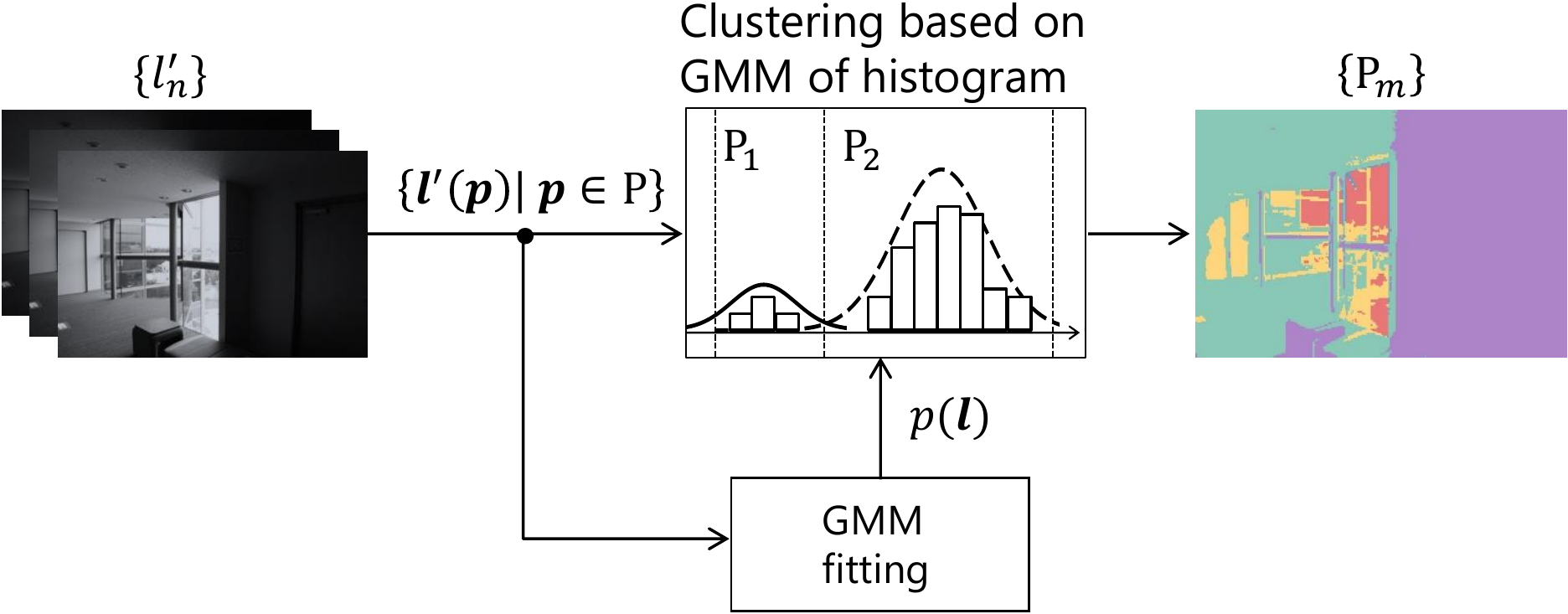}
      \caption{Scene segmentation (Approach 2) \label{fig:ssls_approach2}}
    \end{subfigure}\\
    \begin{subfigure}[t]{\hsize}
      \centering
      \includegraphics[clip, width=\columnwidth]{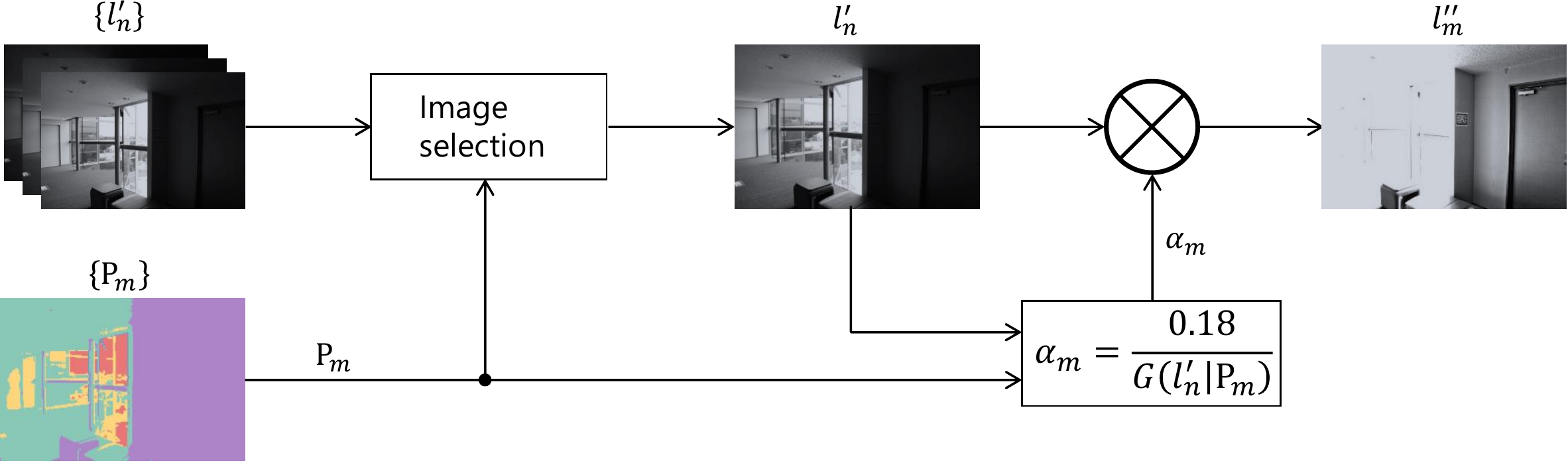}
      \caption{Luminance scaling \label{fig:ssls_scaling}}
    \end{subfigure}
    \caption{Proposed scene segmentation-based luminance scaling.
      (\subref{fig:ssls_flow}) Overview flowchart of SSLS.
      (\subref{fig:ssls_approach1}) and (\subref{fig:ssls_approach2})
      Two approaches to scene segmentation in Fig. \ref{fig:ssls}(\subref{fig:ssls_flow}).
      (\subref{fig:ssls_scaling})
      Luminance scaling in Fig. \ref{fig:ssls}(\subref{fig:ssls_flow}).
      \label{fig:ssls}}
  \end{figure}

  The proposed SSLS is applied to the set $\{l'_n\}$ after
  local contrast enhancement as shown in Fig. \ref{fig:proposedMEF}.
  The purpose of SSLS is to produce luminance $l''_m$,
  which clearly represents area $\myset{P}_m \subset \myset{P}$
  having a specific brightness range in a scene.
  For this purpose, we introduce two processes: scene segmentation and luminance scaling
  (see Fig. \ref{fig:ssls}).
  In the former, the set $\myset{P}$ of all pixels is divided into $M$ subsets
  $\{\myset{P}_1, \myset{P}_2, \cdots, \myset{P}_M\}$.
  A set $\{l''_m\}$ of luminance is produced in the latter process,
  so that luminance $l''_m$ clearly represents area $\myset{P}_m$.
  In the latter process, luminance scaling is applied to all luminance values $l'_n(p)$
  on $\myset{P}$, not on $\myset{P}_m$.
  This enables us to have the same image structure as before the scaling.
\subsubsection{Scene segmentation}
  Here,
  we discuss how to separate a scene in multi-exposure images
  into areas $\myset{P}_1, \cdots, \myset{P}_M$,
  where each has a specific brightness
  and $\myset{P}_1 \cup \myset{P}_2 \cup \cdots \cup \myset{P}_M = \myset{P}$.
  This operation is regarded as image segmentation based on luminance values.
  This segmentation for multi-exposure image fusion
  differs from typical segmentation problems such as semantic segmentation
  \cite{kanezaki2018unsupervised, chen2018deeplab}
  in two ways.
  \begin{itemize}[nosep]
    \item Multiple input images are given.
    \item Attention to the structure of images, e.g., edges, is not needed.
  \end{itemize}
  We propose two approaches to segmentation.
  The first approach (Approach 1), which has low computational cost,
  separates a scene in multi-exposure images,
  according to the luminance values of an input image having middle brightness.
  In the second approach (Approach 2), which provides high-quality results,
  the scene is separated by considering the luminance distribution of all input images.

  \textbf{Approach 1:}
  Let $l'_{\mathrm{med}}$ be the luminance having the middle brightness
  in the set $\{l'_n\}$.
  The overexposed (or underexposed) areas in $l'_{\mathrm{med}}$ are smaller than
  those in the other luminance images in $\{l'_n\}$.
  Namely, $l'_{\mathrm{med}}$ has the best quality in $\{l'_n\}$.
  We thus utilize $l'_{\mathrm{med}}$ to separate a scene.
  
  In Approach 1, $\myset{P}_1, \cdots, \myset{P}_M$ are given by dividing
  the luminance range of $l'_{\mathrm{med}}$ into $M$ equal parts:
  \begin{equation}
    \myset{P}_m
      = \{\myvector{p}|\theta_m \le l'_{\mathrm{med}}(\myvector{p}) \le \theta_{m+1}\},
    \label{eq:luminance_part}
  \end{equation}
  where $\theta_m$ is calculated as
  \begin{align}
    \theta_m & = \frac{N-m+1}{N}
               \left(\max l'_{\mathrm{med}}(\myvector{p})
               - \min l'_{\mathrm{med}}(\myvector{p})\right) \nonumber \\
             & + \min l'_{\mathrm{med}}(\myvector{p}).
    \label{eq:threshold}
  \end{align}
  
  Approach 1 is very simple but is effective for most of the images shown later.

  \textbf{Approach 2:}
  Approach 2 takes into account the luminance distribution of all input images,
  while Approach 1 only uses one of the input images for segmentation.
  This enables us to produce a high-quality fused image, $\myvector{y}$, even when
  Approach 1 does not work well.
  A Gaussian mixture distribution is utilized to model
  the luminance distribution of the input images.
  After that, pixels are classified by using a clustering algorithm based on
  a Gaussian mixture model (GMM) \cite{bishop2006pattern}.

  To obtain a model considering the luminance values of all input images,
  we regard luminance values at a pixel $\myvector{p}$ as an $N$-dimensional vector
  $\myvector{l}' (\myvector{p})
  = (l'_1 (\myvector{p}), l'_2 (\myvector{p}), \cdots, l'_N (\myvector{p}))^\top$.
  Then the distribution of $\myvector{l}' (\myvector{p})$ is modeled by using a GMM as
  \begin{equation}
    p(\myvector{l'} (\myvector{p})) =
      \sum_{k=1}^K \pi_k
      \gaussian{\myvector{l}' (\myvector{p})}{\myvector{\mu}_k}{\mymatrix{\Sigma}_k},
    \label{eq:gmm}
  \end{equation}
  where $K$ is the number of mixture components, $\pi_k$ is a mixing coefficient, and
  $\gaussian{\myvector{l}' (\myvector{p})}{\myvector{\mu}_k}{\myvector{\Sigma}_k})$
  is an $N$-dimensional Gaussian distribution
  with mean $\myvector{\mu}_k$ and variance-covariance matrix $\mymatrix{\Sigma}_k$.
  
  To fit the GMM into a given $\myvector{l}' (\myvector{p})$,
  variational Bayesian inference \cite{bishop2006pattern} is utilized.
  Compared with the traditional maximum likelihood estimation,
  one of the advantages is that variational Bayesian inference can avoid overfitting
  even when we choose a large $K$.
  For this reason, unnecessary mixture components are automatically removed by using
  the inference together with a large $K$.
  $K = 10$ is used in this paper as the maximum of the partition number $M$.

  Let us introduce a $K$-dimensional binary random variable, $\myvector{z}$, having a 1-of-$K$
  representation in which a particular element $z_k$ is equal to 1 and all other elements are
  equal to 0.
  The values of $z_k$ therefore satisfy $z_k \in \{0, 1\}$ and $\sum_k z_k = 1$.
  The marginal distribution over $\myvector{z}$ is specified in terms of
  a mixing coefficient, $\pi_k$, such that
  \begin{equation}
    p(z_k = 1) = \pi_k.
  \end{equation}
  In order for $p(z_k = 1)$ to be a valid probability,
  $\{\pi_k\}$ must satisfy
  \begin{equation}
    0 \leq \pi_k \leq 1
  \end{equation}
  together with
  \begin{equation}
    \sum_{k=1}^K \pi_k = 1.
  \end{equation}
  An area $\myset{P}_m$ containing a pixel $\myvector{p}$ is determined by
  the responsibility $\gamma (z_k | \myvector{l}' (\myvector{p}))$,
  which is given as the conditional probability:
  \begin{align}
    \gamma (z_k | \myvector{l}' (\myvector{p}))
      &= p(z_k = 1 | \myvector{l}' (\myvector{p})) \nonumber \\
      &= \frac{
          \pi_k\gaussian{\myvector{l}' (\myvector{p})}{\myvector{\mu}_k}{\mymatrix{\Sigma}_k}}
        {\sum_{j=1}^K \pi_j\gaussian{\myvector{l}' (\myvector{p})}
          {\myvector{\mu}_j}{\mymatrix{\Sigma}_j}}.
      \label{eq:responsibility}
  \end{align}
  When a pixel $\myvector{p} \in \myset{P}$ is given and $m$ satisfies
  \begin{equation}
    m = \argmax_k \gamma (z_k | \myvector{l}' (\myvector{p})),
    \label{eq:classLabel}
  \end{equation}
  the pixel $\myvector{p}$ is assigned to a subset $\myset{P}_m$ of $\myset{P}$.
\subsubsection{\label{sec:luminance_scaling} Luminance scaling}
  The scaled luminance $l''_m$ is simply obtained by
  \begin{equation}
    l''_m (\myvector{p}) = \alpha_m l'_n (\myvector{p}),
    \label{eq:constMultiplication}
  \end{equation}
  where parameter $\alpha_m > 0$ indicates the degree of adjustment
  (see Appendix \ref{sec:relationship}).

  Given $\myset{P}_m$ as a subset of $\myset{P}$,
  the approximate brightness of $\myset{P}_m$ is calculated
  as the geometric mean of luminance values on $\myset{P}_m$.
  We thus estimate an adjusted multi-exposure image $l''_m (\myvector{p})$
  so that the geometric mean of its luminance equals the middle-gray
  of the displayed image, or 0.18 on a scale from zero to one,
  as in \cite{reinhard2002photographic}.
  
  The geometric mean $G(l|\myset{P}_m)$ of luminance $l$
  on pixel set $\myset{P}_m$ is calculated by using
  \begin{equation}
    G(l|\myset{P}_m) =
      \exp{
        \left(\frac{1}{|\myset{P}_m|}
          \sum_{\myvector{p} \in \myset{P}_m}
          \log{\left(\max{\left( l(\myvector{p}), \epsilon \right)}\right)}
        \right)
      },
    \label{eq:geoMeanEps}
  \end{equation}
  where $\epsilon$ is set to a small value to avoid singularities at $l(\myvector{p})=0$.

  By using eq. (\ref{eq:geoMeanEps}),
  parameter $\alpha_m$ is calculated as
  \begin{equation}
    \alpha_m = \frac{0.18}{G(l'_n|\myset{P}_{m})}.
    \label{eq:unknownEV}
  \end{equation}
  Since a smaller value for parameter $\alpha_m$ is better,
  $n$ is chosen as
  \begin{equation}
    n = \psi(m) = \argmin_{j} (0.18 - G(l'_j | \myset{P}_{m}))^2.
    \label{eq:associating}
  \end{equation}
  The scaled luminance $l''_m$, calculated by using eq. (\ref{eq:constMultiplication})
  with parameters $\alpha_m$ and $n$, is used as input of the tone mapping operation
  described in \ref{sec:tone_mapping}.
\subsection{Combining adjusted luminance and input images}
  Adjusted images $\{\hat{\myvector{x}}_m\}$ are obtained by combining
  a set $\{\hat{l}_m\}$ of luminance adjusted by the SSLA and
  input multi-exposure images $\{\myvector{x}_n\}$.
  To associate each $\hat{l}_m$ with an input image $\myvector{x}_n$,
  eq. (\ref{eq:associating}) is utilized.
  As a result,
  combining $\hat{l}_m$, the $\psi(m)$-th input image $\myvector{x}_{\psi(m)}$,
  and its luminance $l_{\psi(m)}$,
  we obtain adjusted multi-exposure images $\hat{\myvector{x}}_m$:
  \begin{equation}
    \hat{\myvector{x}}_m (\myvector{p})
      = \frac{\hat{l}_m (\myvector{p})}{l_{\psi(m)} (\myvector{p})}
      \myvector{x}_{\psi(m)} (\myvector{p}).
    \label{eq:color}
  \end{equation}
  This transformation can preserve original colors
  since the RGB color space has a linear relationship with the XYZ color space
  (see Appendix \ref{sec:color}).
\subsection{Proposed procedure}
  The procedure for generating an image $\myvector{y}$ from
  $N$ input images $\{\myvector{x}_n\}$
  with the proposed method is summarized as
  follows (see Fig. \ref{fig:proposedMEF}).
  \begin{enumerate}[label=\roman*, align=parleft, leftmargin=*]
    \item Calculate the luminance $l_n$ from each input image $x_n$.
    \item Enhance the local contrast of ${l_n}$ by using eq. (\ref{eq:dodgingAndBurning})
      to eq. (\ref{eq:gaussian})
      and then obtain the enhanced luminance $\{l'_n\}$.
    \item Separate $\myset{P}$ into $M$ areas $\{\myset{P}_m\}$
      \begin{enumerate}[label=Approach \arabic*:, align=parleft, leftmargin=*]
        \item use eqs. (\ref{eq:luminance_part}) and (\ref{eq:threshold}).
        \item use eq. (\ref{eq:gmm}) to eq. (\ref{eq:classLabel}).
      \end{enumerate}
    \item Calculate $\{l''_m\}$ by using
      eq. (\ref{eq:constMultiplication}) to eq. (\ref{eq:associating}).
    \item Map $\{l''_m\}$ to $\{\hat{l}_m\}$ according to
      eqs. (\ref{eq:TM}) and (\ref{eq:reinhardTMO}).
    \item Generate $\{\hat{\myvector{x}}_m\}$ according to eq. (\ref{eq:color}).
    \item Obtain an image $\myvector{y}$ with
      a multi-exposure image fusion method $\mathscr{F}$
      as in eq. (\ref{eq:fusion}).
  \end{enumerate}
  Note that the difference between Approach 1 and Approach 2 is only in
  step \rnum{3}, and the number $M$ of $\myset{P}_m$ satisfies $1 \le M \le K$.
  Tunable parameters in the proposed method are shown in Table \ref{tab:parameters}.
\begin{table}[!t]
  \centering
  \caption{Tunable parameters in proposed method}
  \begin{tabular}{l|p{7cm}}\hline\hline
    Parameter & \multicolumn{1}{|c}{Effect} \\ \hline
    $\sigma_1$ and $\sigma_2$ & Determine the degree of enhancing local contrast,
      but larger values often boost noise or ringing.\\
    $L_m$ & Determine the shape of tone mapping operators: $L_m \le 1$ increases contrast,
      $L_m = 1$ keeps all luminance values, and $1 \le L_m$ suppresses contrast.\\
    $M$ and $K$ & Determine the number of adjusted multi-exposure images.
      Larger $M$ improves the quality of fused images, but increases computational cost.
      In Approach 2, $M$ is automatically set such that $1 \le M \le K$.\\
    \hline
  \end{tabular}
  \label{tab:parameters}
\end{table}

\section{Simulation}
  We evaluated the effectiveness of the proposed SSLA
  in terms of the quality of fused images $\myvector{y}$.
\subsection{Simulation conditions}
\begin{figure*}[!t]
  \centering
  \begin{subfigure}[t]{0.25\hsize}
    \centering
    \includegraphics[width=\columnwidth]
      {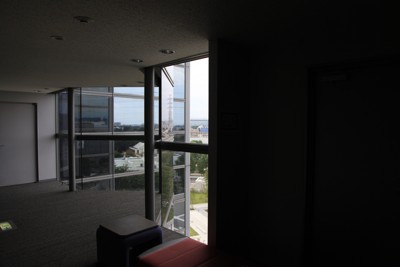}
    \caption{Input image $\myvector{x}_1$ ($-1 \mathrm{[EV]}$)}
  \end{subfigure}
  \begin{subfigure}[t]{0.25\hsize}
    \includegraphics[width=\columnwidth]
      {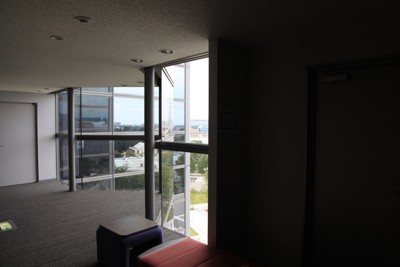}
    \caption{Input image $\myvector{x}_2$ ($0 \mathrm{[EV]}$)}
  \end{subfigure}
  \begin{subfigure}[t]{0.25\hsize}
    \includegraphics[width=\columnwidth]
      {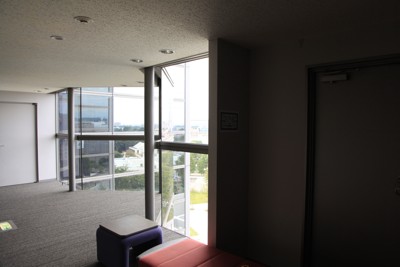}
    \caption{Input image $\myvector{x}_3$ ($+1 \mathrm{[EV]}$)}
  \end{subfigure}
  \caption{Input multi-exposure images for ``Window.''
    Right area in scene is under-exposed in all images \label{fig:input_window}}
\end{figure*}

\begin{figure*}[!t]
  \centering
  \begin{subfigure}[t]{0.25\hsize}
    \centering
    \includegraphics[width=\columnwidth]
      {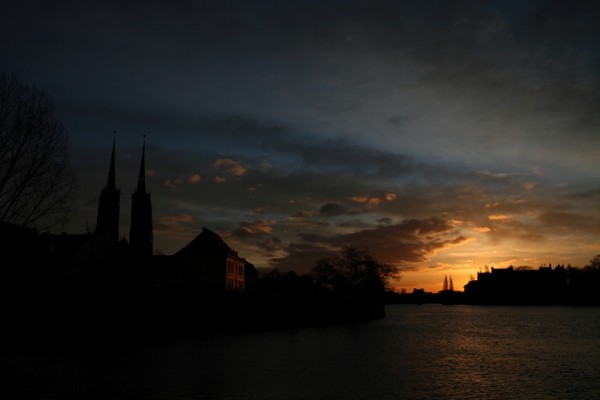}
    \caption{Input image $\myvector{x}_1$ ($-2 \mathrm{[EV]}$)}
  \end{subfigure}
  \begin{subfigure}[t]{0.25\hsize}
    \includegraphics[width=\columnwidth]
      {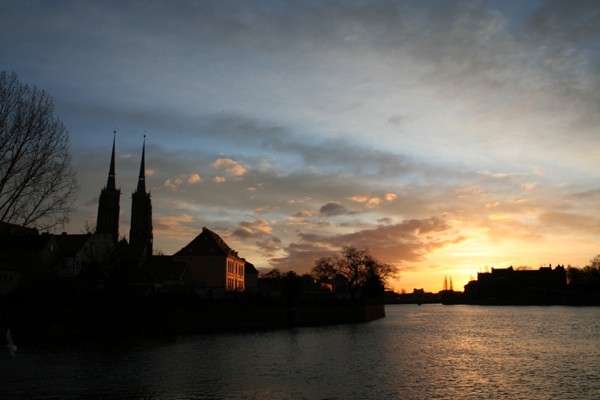}
    \caption{Input image $\myvector{x}_2$ ($0 \mathrm{[EV]}$)}
  \end{subfigure}
  \begin{subfigure}[t]{0.25\hsize}
    \includegraphics[width=\columnwidth]
      {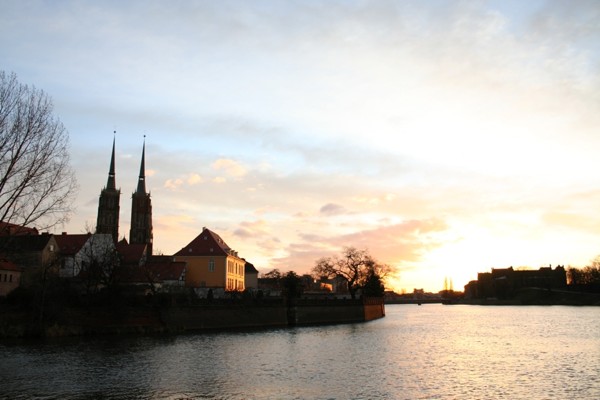}
    \caption{Input image $\myvector{x}_3$ ($+2 \mathrm{[EV]}$)}
  \end{subfigure}
  \caption{Input multi-exposure images for ``Ostrow Tumski.''
    Towers and land on right side are under-exposed in all images.
    \label{fig:input_ostrow}}
\end{figure*}

  In the simulation,
  20 sets of photographs and 550 sets of
  tone-mapped images were used
  as input images $\{\myvector{x}_n\}$.
  Among the sets of photographs,
  four were taken with a Canon EOS 5D Mark II camera,
  and eight were selected from an available online database \cite{easyhdr},
  where each set contained three multi-exposure images that were exposed with
  negative, zero, and positive exposure values.
  (see Figs. \ref{fig:input_window} and \ref{fig:input_ostrow}).
  The other sets were collected from database \cite{sen2012robust},
  where each set included moving objects.
  The sets of tone-mapped images were generated by tone-mapping
  50 HDR images selected from online databases \cite{zolliker2013creating, hdrps}.
  From each HDR image, 11 sets of images were generated
  by linear response functions,
  where the number of images in ten sets was randomly
  decided in the range of $[2, 5]$
  and their exposure values were also randomly
  determined in the range of $[-7, 0]$,
  to produce unclear multi-exposure images.
  The other set contained $15$ multi-exposure images
  whose exposure values were $-7, -6, \cdots, 6$, and $7$.

  The following procedure was carried out to evaluate the effectiveness of the proposed method.
  \begin{enumerate}[label=\roman*, align=parleft, leftmargin=*]
    \item Produce $\{\hat{\myvector{x}}_m\}$
      from $\{\myvector{x}_n\}$ by using the proposed method.
    \item Obtain $\myvector{y}$ fused from $\{\hat{\myvector{x}}_m\}$ with $\mathscr{F}$.
    \item Compute scores of three objective quality metrics: MEF-SSIM,
      discrete entropy and TMQI,
      for $\myvector{y}$, as described later.
  \end{enumerate}
  Here, we used five fusion methods for $\mathscr{F}$:
  Mertens's method \cite{mertens2009exposure},
  Sakai's method \cite{sakai2015hybrid},
  Nejati's method \cite{nejati2017fast},
  Li's method \cite{li2013image},
  and Ma's method \cite{ma2017robust}.
  In Approach 1, $M$ was set so as to be the same as the number $N$ of input images.
  In Approach 2, $M$ was a number that satisfies $1 \le M \le K = 10$,
  and the number $M$ was determined by the resulting GMM for each scene.
  The fitting of a GMM was cut off when the number of iterations reached 100,
  regardless of parameter convergence.
  In addition, Approach 2 was applied to downsized versions of
  $\{l'_n\}$ for fast calculation.
  In particular, the resized width and height ($U'$ and $V'$, respectively) of $l'_n$
  were determined so that $\max (U', V') = 256$.
\subsection{Visual evaluation}
\begin{figure*}[!t]
  \centering
  \begin{subfigure}[t]{0.31\hsize}
    \centering
    \includegraphics[width=\columnwidth]
      {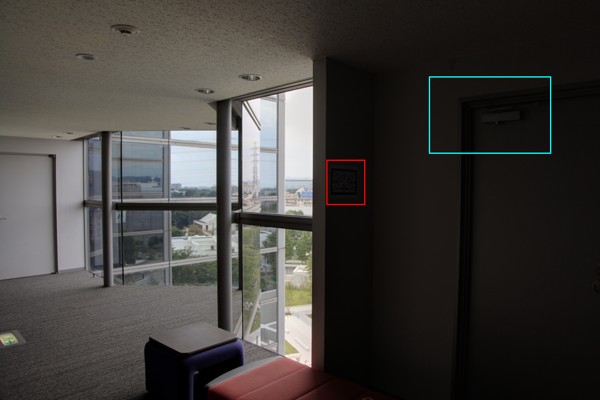}\\
    \includegraphics[width=0.34\columnwidth]
      {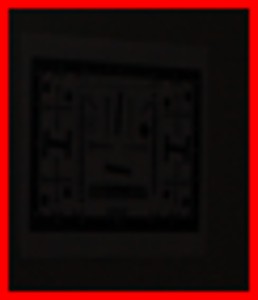}
    \includegraphics[width=0.63\columnwidth]
      {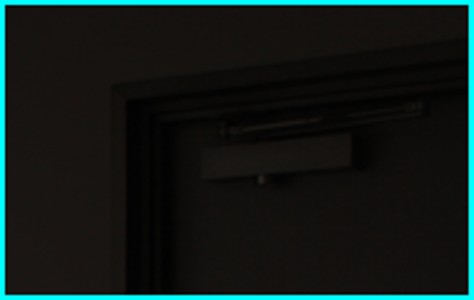}
    \caption{Mertens {\protect\cite{mertens2009exposure}}.
      Entropy: 5.047\\
      and Naturalness: 0.0569.
      \label{fig:fused_window_org_mertens}}
  \end{subfigure}
  \begin{subfigure}[t]{0.31\hsize}
    \centering
    \includegraphics[width=\columnwidth]
      {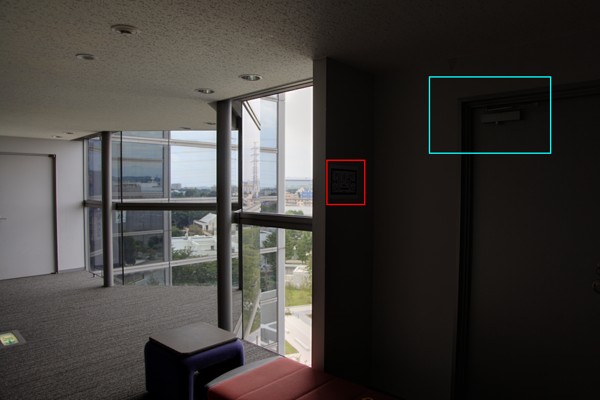}\\
    \includegraphics[width=0.34\columnwidth]
      {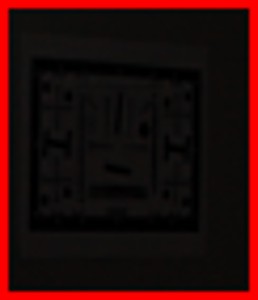}
    \includegraphics[width=0.63\columnwidth]
      {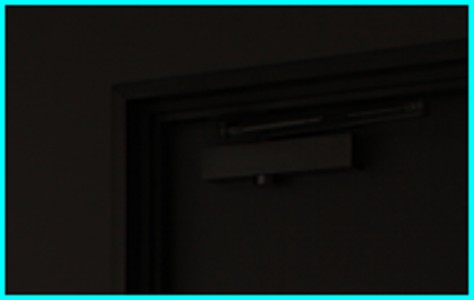}
    \caption{Sakai {\protect\cite{sakai2015hybrid}}.
      Entropy: 5.052\\
      and Naturalness: 0.0615.
      \label{fig:fused_window_org_sakai}}
  \end{subfigure}
  \begin{subfigure}[t]{0.31\hsize}
    \centering
    \includegraphics[width=\columnwidth]
      {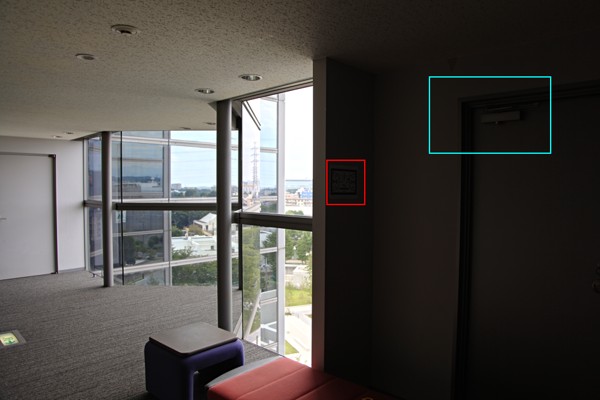}\\
    \includegraphics[width=0.34\columnwidth]
      {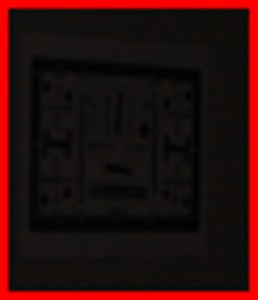}
    \includegraphics[width=0.63\columnwidth]
      {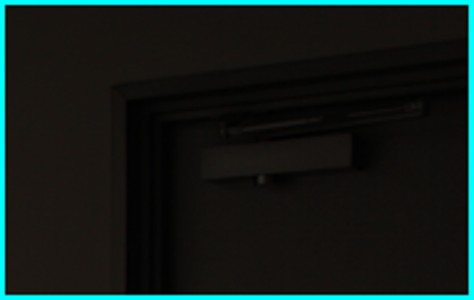}
    \caption{Nejati {\protect\cite{nejati2017fast}}.
      Entropy: 5.407\\
      and Naturalness: 0.1617.
      \label{fig:fused_window_org_nejati}}
  \end{subfigure}\\
  \begin{subfigure}[t]{0.31\hsize}
    \centering
    \includegraphics[width=\columnwidth]
      {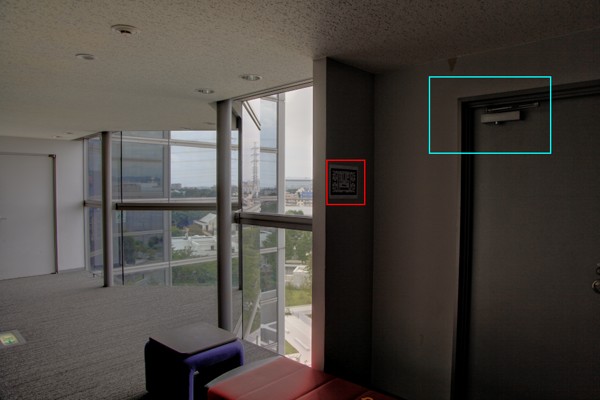}\\
    \includegraphics[width=0.34\columnwidth]
      {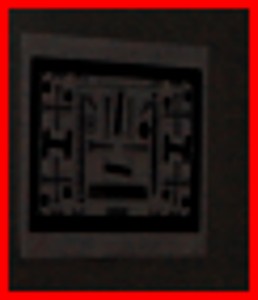}
    \includegraphics[width=0.63\columnwidth]
      {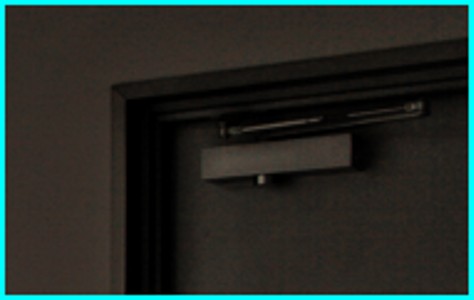}
    \caption{Propoced method with Mertens (Approach 1).
      Entropy: 5.715
      and Naturalness: 0.0912.
      \label{fig:fused_window_range_mertens}}
  \end{subfigure}
  \begin{subfigure}[t]{0.31\hsize}
    \centering
    \includegraphics[width=\columnwidth]
      {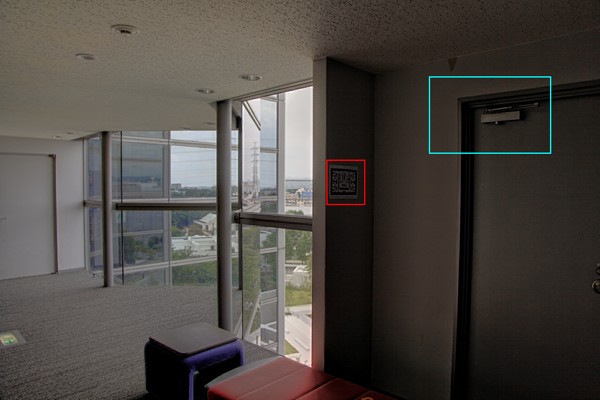}\\
    \includegraphics[width=0.34\columnwidth]
      {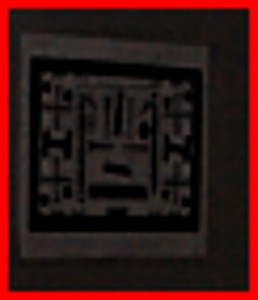}
    \includegraphics[width=0.63\columnwidth]
      {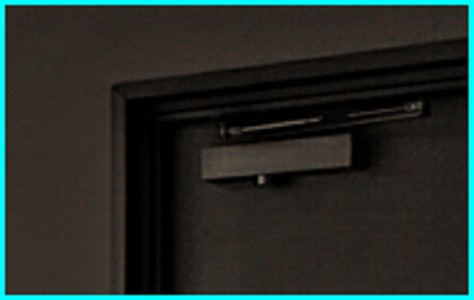}
    \caption{Proposed method with Sakai (Approach 1).
      Entropy: 5.735
      and Naturalness: 0.1142.
      \label{fig:fused_window_range_sakai}}
  \end{subfigure}
  \begin{subfigure}[t]{0.31\hsize}
    \centering
    \includegraphics[width=\columnwidth]
      {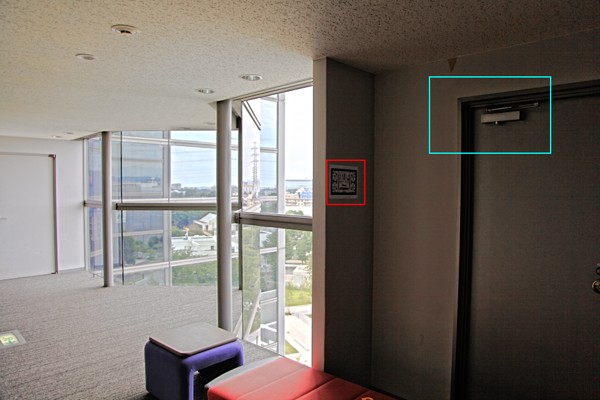}\\
    \includegraphics[width=0.34\columnwidth]
      {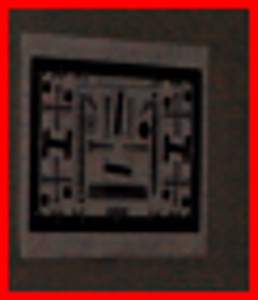}
    \includegraphics[width=0.63\columnwidth]
      {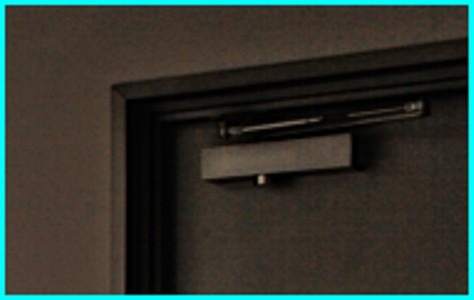}
    \caption{Proposed method with Nejati (Approach 1).
      Entropy: 6.750
      and Naturalness: 0.6187.
      \label{fig:fused_window_range_nejati}}
  \end{subfigure}\\
  \begin{subfigure}[t]{0.31\hsize}
    \centering
    \includegraphics[width=\columnwidth]
      {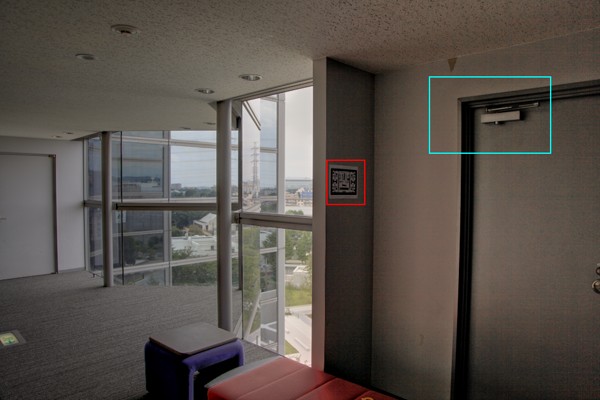}\\
    \includegraphics[width=0.34\columnwidth]
      {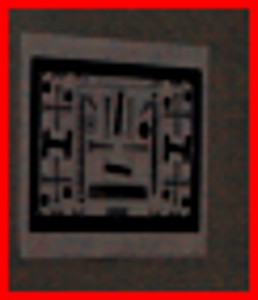}
    \includegraphics[width=0.63\columnwidth]
      {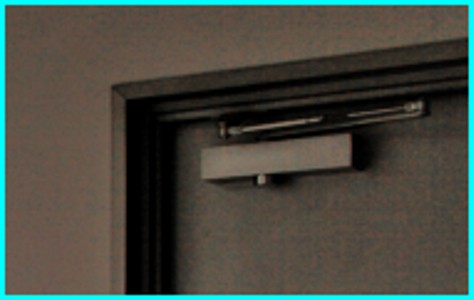}
    \caption{Proposed method with Mertens (Approach 2).
      Entropy: 6.060
      and Naturalness: 0.2135.
      \label{fig:fused_window_variational_mertens}}
  \end{subfigure}
  \begin{subfigure}[t]{0.31\hsize}
    \centering
    \includegraphics[width=\columnwidth]
      {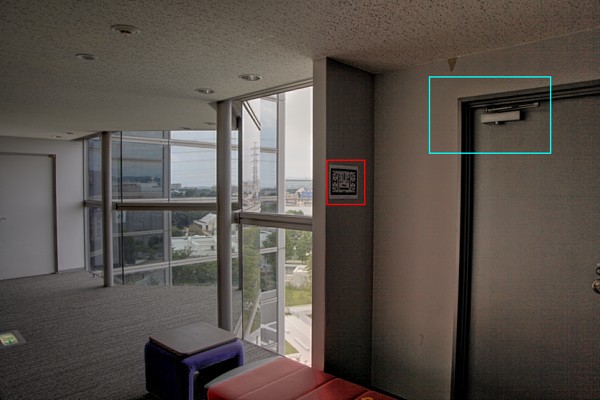}\\
    \includegraphics[width=0.34\columnwidth]
      {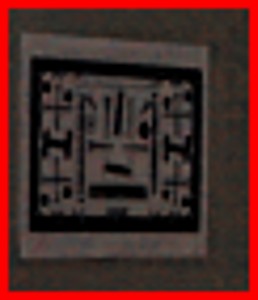}
    \includegraphics[width=0.63\columnwidth]
      {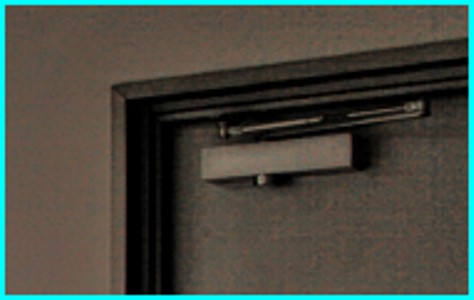}
    \caption{Proposed method with Sakai (Approach 2).
      Entropy: 6.072
      and Naturalness: 0.2650.
      \label{fig:fused_window_variational_sakai}}
  \end{subfigure}
  \begin{subfigure}[t]{0.31\hsize}
    \centering
    \includegraphics[width=\columnwidth]
      {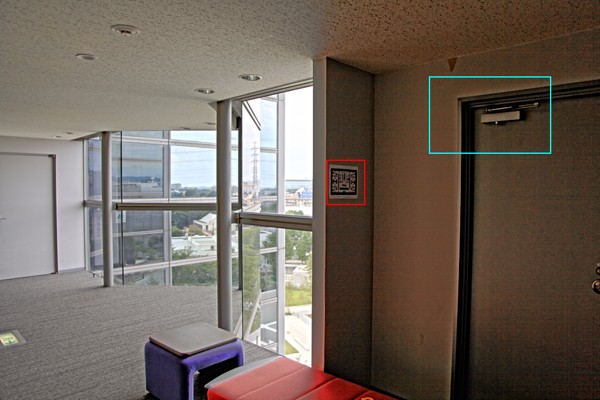}\\
    \includegraphics[width=0.34\columnwidth]
      {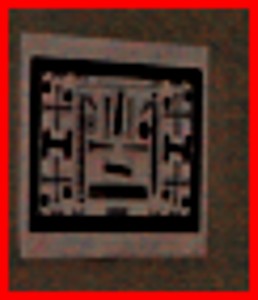}
    \includegraphics[width=0.63\columnwidth]
      {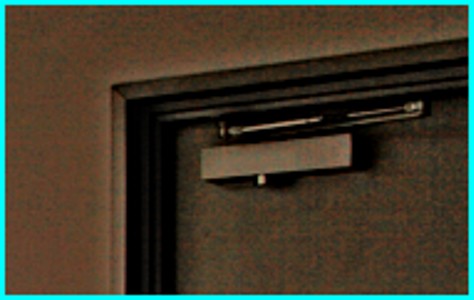}
    \caption{Proposed method with Nejati (Approach 2).
      Entropy: 6.848
      and Naturalness: 0.7031.
      \label{fig:fused_window_variational_nejati}}
  \end{subfigure}
  \caption{Comparison of proposed method with Mertens's, Sakai's, and Nejati's
    methods (``Window'').
    Zoom-ins of boxed regions are shown in bottom of each fused image.
    Conventional fusion methods without adjustment do not produce clear images
    from unclear multi-exposure images shown in Fig. {\protect\ref{fig:input_window}}.
    Proposed method enables us to produce clear images.
    \label{fig:fused_window}}
\end{figure*}

\begin{figure*}[!t]
  \centering
  \begin{subfigure}[t]{0.31\hsize}
    \centering
    \includegraphics[width=\columnwidth]
      {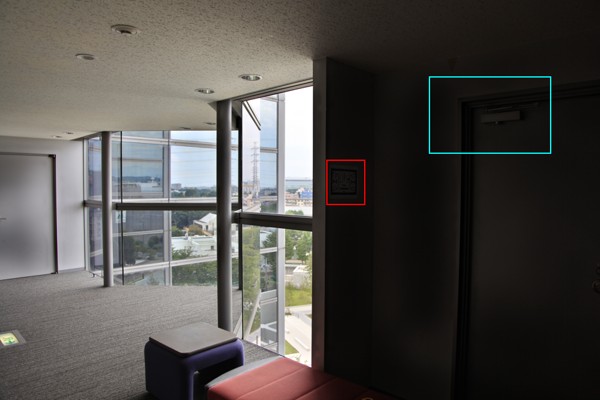}\\
    \includegraphics[width=0.34\columnwidth]
      {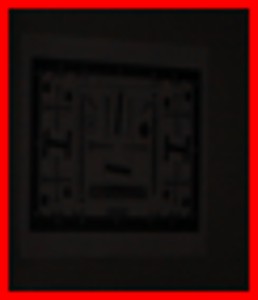}
    \includegraphics[width=0.63\columnwidth]
      {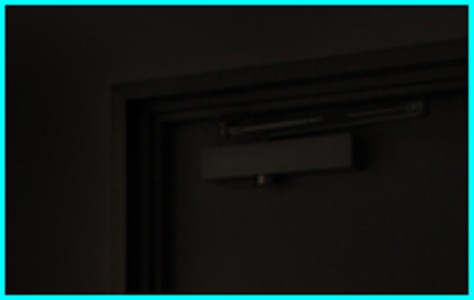}
    \caption{Li {\protect\cite{li2013image}}.
      Entropy: 5.547\\
      and Naturalness: 0.2145.
      \label{fig:fused_window_org_li}}
  \end{subfigure}
  \begin{subfigure}[t]{0.31\hsize}
    \centering
    \includegraphics[width=\columnwidth]
      {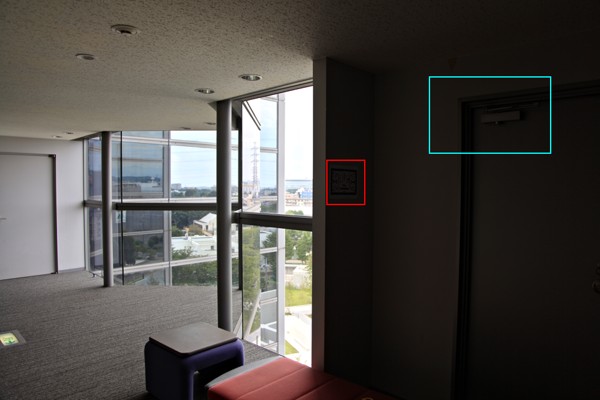}\\
    \includegraphics[width=0.34\columnwidth]
      {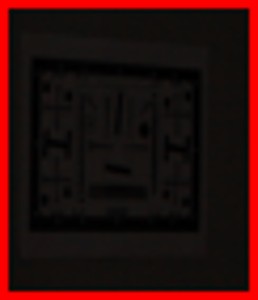}
    \includegraphics[width=0.63\columnwidth]
      {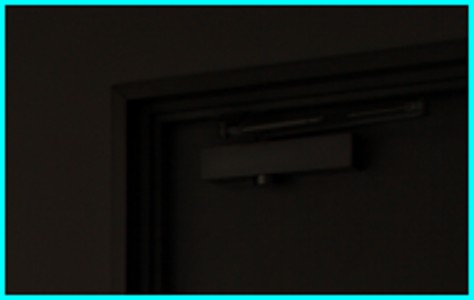}
    \caption{Ma {\protect\cite{ma2017robust}}.
      Entropy: 5.402\\
      and Naturalness: 0.1655.
      \label{fig:fused_window_org_ma}}
  \end{subfigure}
  \begin{subfigure}[t]{0.31\hsize}
    \centering
    \includegraphics[width=\columnwidth]
      {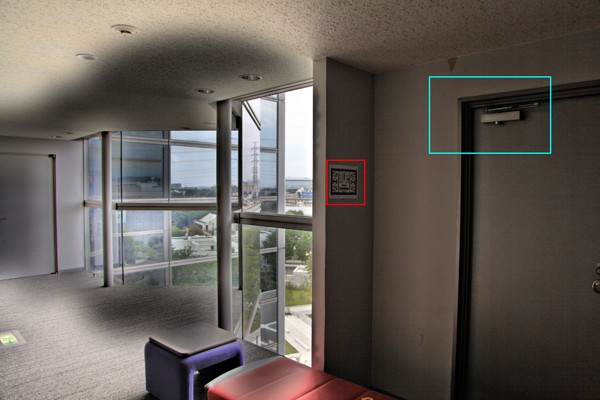}\\
    \includegraphics[width=0.34\columnwidth]
      {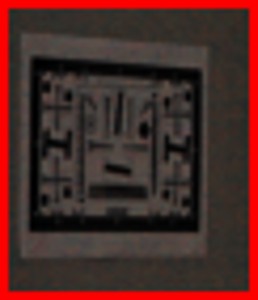}
    \includegraphics[width=0.63\columnwidth]
      {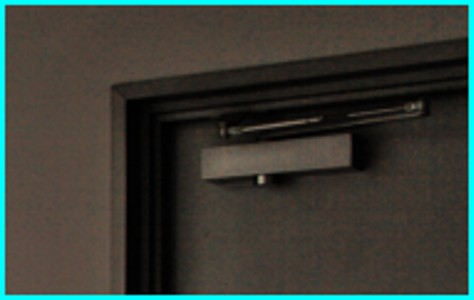}
    \caption{Proposed method with Li (Approach 1).
      Entropy: 6.617
      and Naturalness: 0.5070.
      \label{fig:fused_window_range_li}}
  \end{subfigure}\\
  \begin{subfigure}[t]{0.31\hsize}
    \centering
    \includegraphics[width=\columnwidth]
      {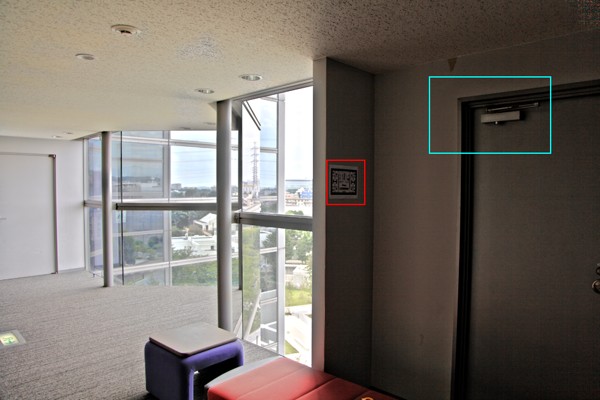}\\
    \includegraphics[width=0.34\columnwidth]
      {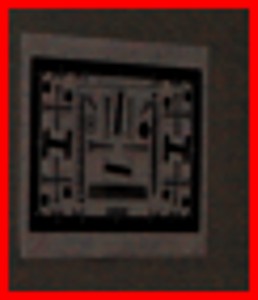}
    \includegraphics[width=0.63\columnwidth]
      {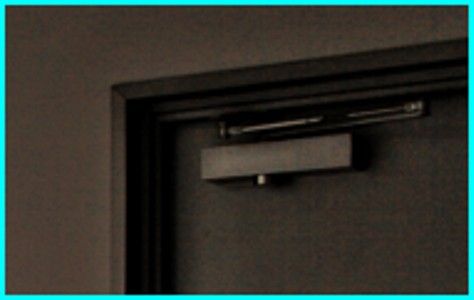}
    \caption{Proposed method with Ma (Approach 1).
      Entropy: 6.752
      and Naturalness: 0.6413.
      \label{fig:fused_window_range_ma}}
  \end{subfigure}
  \begin{subfigure}[t]{0.31\hsize}
    \centering
    \includegraphics[width=\columnwidth]
      {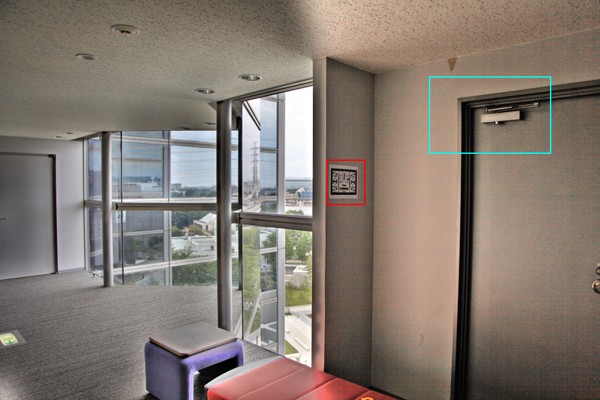}\\
    \includegraphics[width=0.34\columnwidth]
      {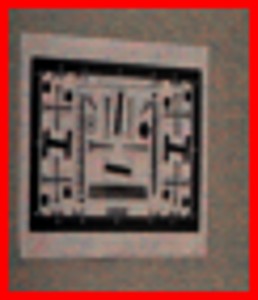}
    \includegraphics[width=0.63\columnwidth]
      {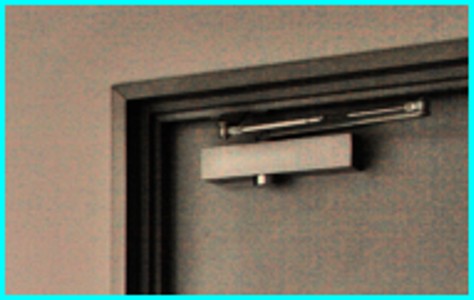}
    \caption{Proposed method with Li (Approach 2).
      Entropy: 7.099
      and Naturalness: 0.8770.
      \label{fig:fused_window_variational_li}}
  \end{subfigure}
  \begin{subfigure}[t]{0.31\hsize}
    \centering
    \includegraphics[width=\columnwidth]
      {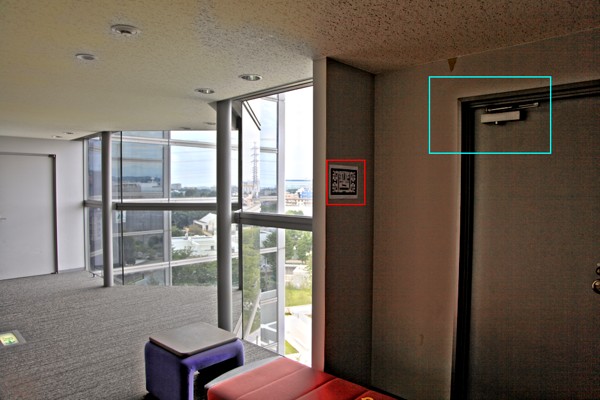}\\
    \includegraphics[width=0.34\columnwidth]
      {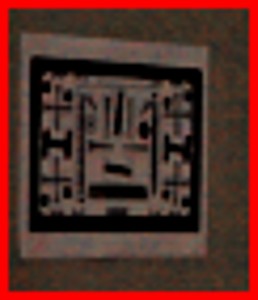}
    \includegraphics[width=0.63\columnwidth]
      {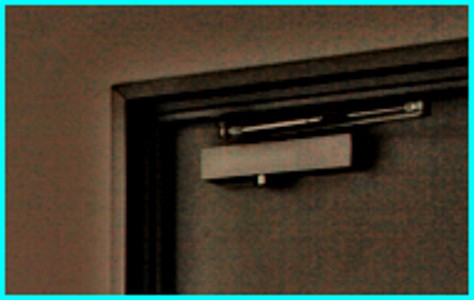}
    \caption{Proposed method with Ma (Approach 2).
      Entropy: 6.868
      and Naturalness: 0.7475.
      \label{fig:fused_window_variational_ma}}
  \end{subfigure}
  \caption{Comparison of proposed method with Li's and Ma's methods (``Window'').
    Zoom-ins of boxed regions are shown in bottom of each fused image.
    Conventional fusion methods without adjustment do not produce clear images
    from unclear multi-exposure images shown in Fig. {\protect\ref{fig:input_window}}.
    Proposed method enables us to produce clear images.
    \label{fig:fused_window_li_ma}}
\end{figure*}

\begin{figure*}[!t]
  \centering
  \begin{subfigure}[t]{0.31\hsize}
    \centering
    \includegraphics[width=\columnwidth]
      {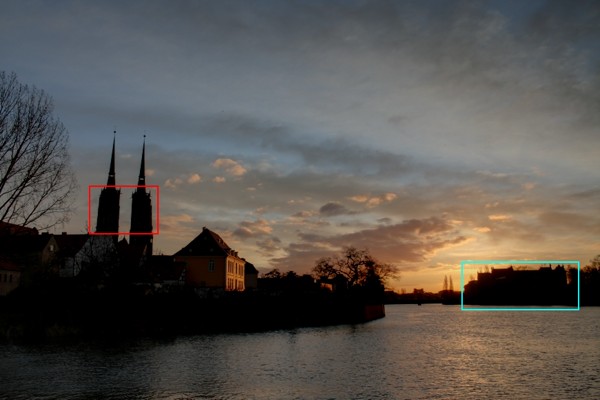}\\
    \includegraphics[width=0.35\columnwidth]
      {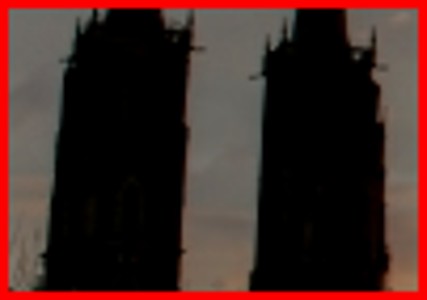}
    \includegraphics[width=0.62\columnwidth]
      {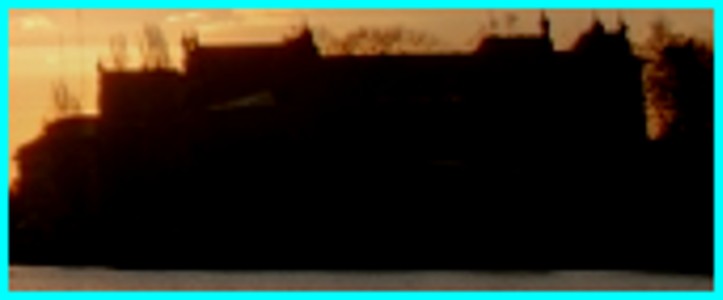}
    \caption{Mertens {\protect\cite{mertens2009exposure}}.
      Entropy: 5.861\\
      and Naturalness: 0.1445.
      \label{fig:fused_ostrow_org_mertens}}
  \end{subfigure}
  \begin{subfigure}[t]{0.31\hsize}
    \centering
    \includegraphics[width=\columnwidth]
      {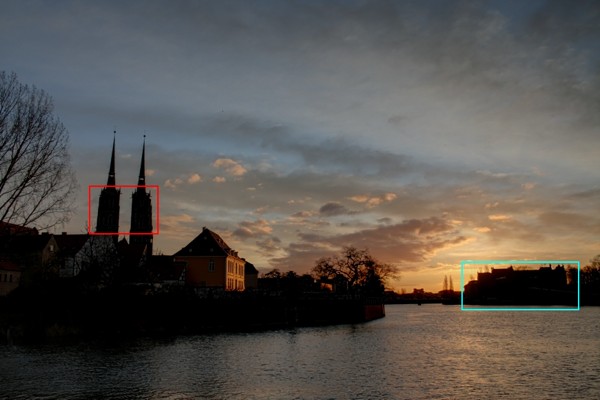}\\
    \includegraphics[width=0.35\columnwidth]
      {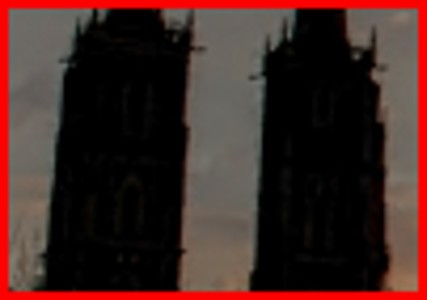}
    \includegraphics[width=0.62\columnwidth]
      {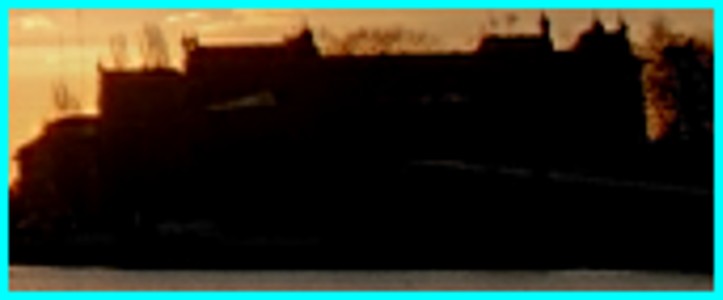}
    \caption{Sakai {\protect\cite{sakai2015hybrid}}.
      Entropy: 5.874\\
      and Naturalness: 0.1580.
      \label{fig:fused_ostrow_org_sakai}}
  \end{subfigure}
  \begin{subfigure}[t]{0.31\hsize}
    \centering
    \includegraphics[width=\columnwidth]
      {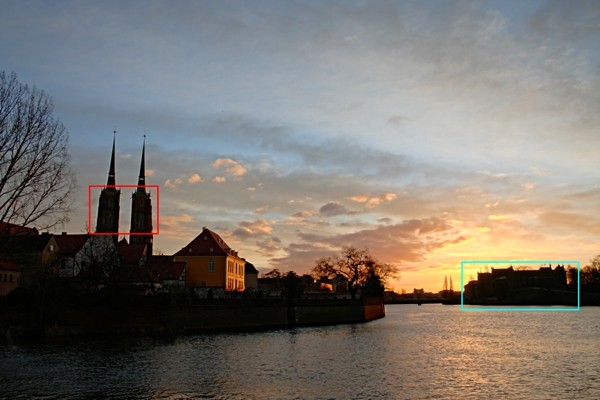}\\
    \includegraphics[width=0.35\columnwidth]
      {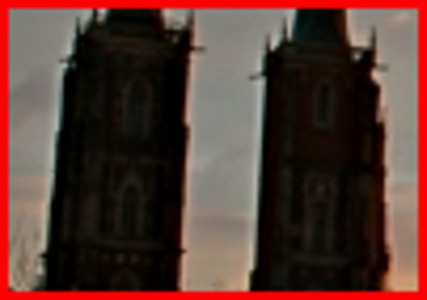}
    \includegraphics[width=0.62\columnwidth]
      {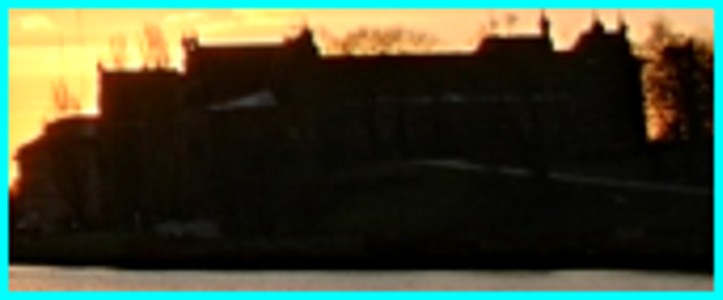}
    \caption{Nejati {\protect\cite{nejati2017fast}}.
      Entropy: 6.812\\
      and Naturalness: 0.5090.
      \label{fig:fused_ostrow_org_nejati}}
  \end{subfigure}\\
  \begin{subfigure}[t]{0.31\hsize}
    \centering
    \includegraphics[width=\columnwidth]
      {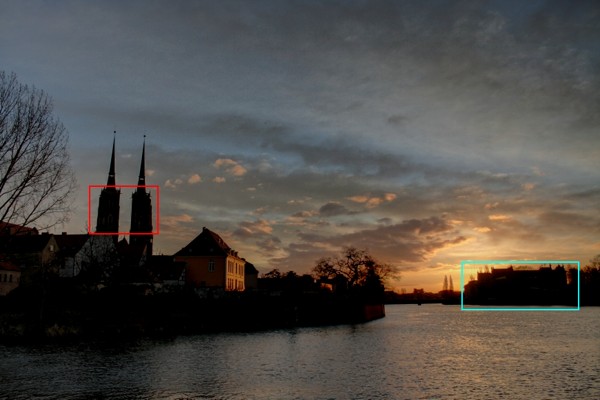}\\
    \includegraphics[width=0.35\columnwidth]
      {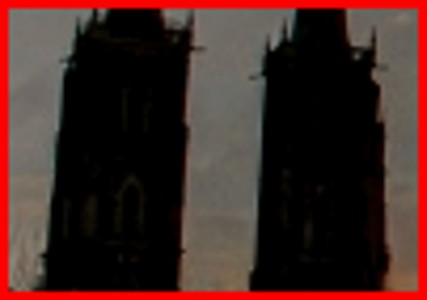}
    \includegraphics[width=0.62\columnwidth]
      {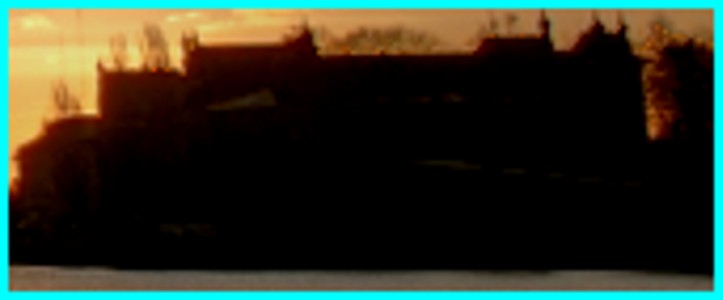}
    \caption{Proposed method with Mertens (Approach 1).
      Entropy: 5.607
      and Naturalness: 0.1347.
      \label{fig:fused_ostrow_range_mertens}}
  \end{subfigure}
  \begin{subfigure}[t]{0.31\hsize}
    \centering
    \includegraphics[width=\columnwidth]
      {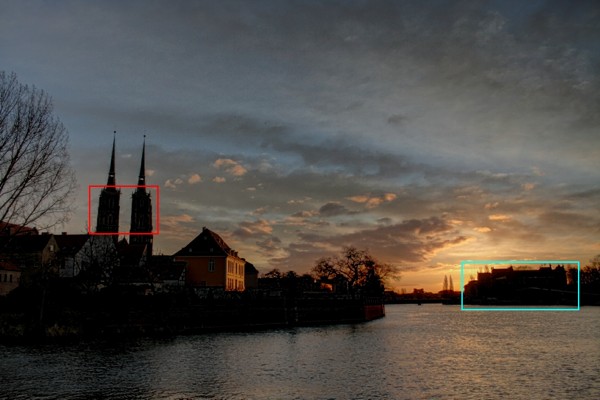}\\
    \includegraphics[width=0.35\columnwidth]
      {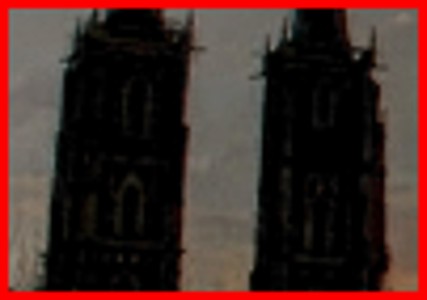}
    \includegraphics[width=0.62\columnwidth]
      {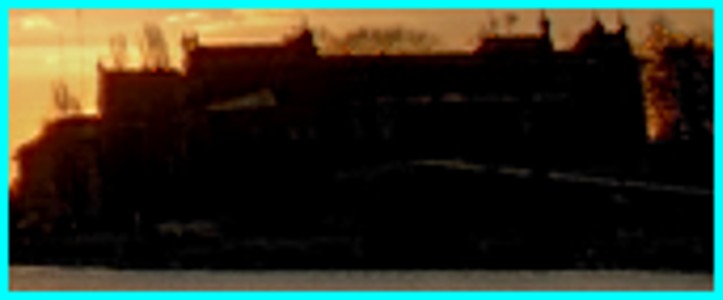}
    \caption{Proposed method with Sakai (Approach 1).
      Entropy: 5.629
      and Naturalness: 0.1516.
      \label{fig:fused_ostrow_range_sakai}}
  \end{subfigure}
  \begin{subfigure}[t]{0.31\hsize}
    \centering
    \includegraphics[width=\columnwidth]
      {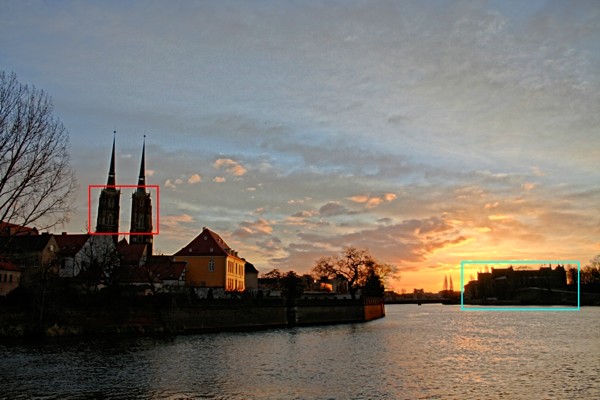}\\
    \includegraphics[width=0.35\columnwidth]
      {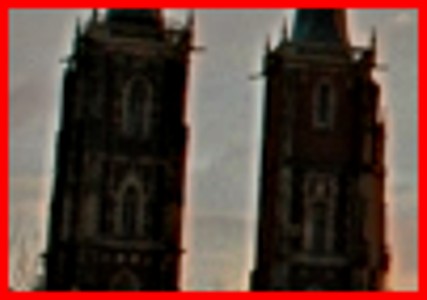}
    \includegraphics[width=0.62\columnwidth]
      {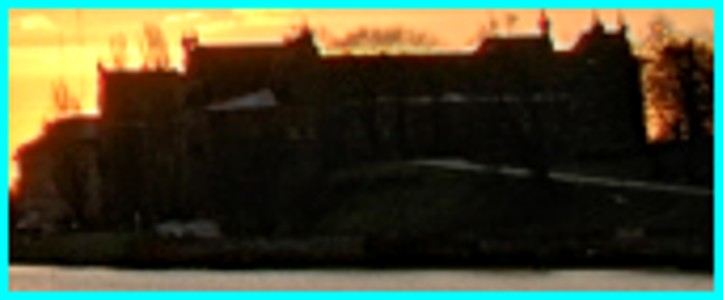}
    \caption{Proposed method with Nejati (Approach 1).
      Entropy: 6.475
      and Naturalness: 0.6194.
      \label{fig:fused_ostrow_range_nejati}}
  \end{subfigure}\\
  \begin{subfigure}[t]{0.31\hsize}
    \centering
    \includegraphics[width=\columnwidth]
      {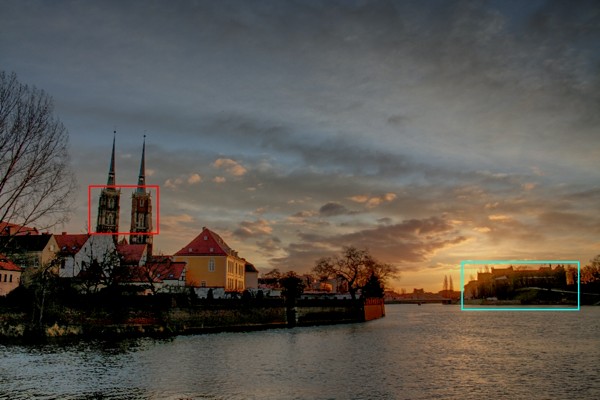}\\
    \includegraphics[width=0.35\columnwidth]
      {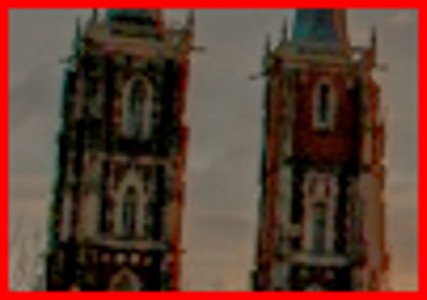}
    \includegraphics[width=0.62\columnwidth]
      {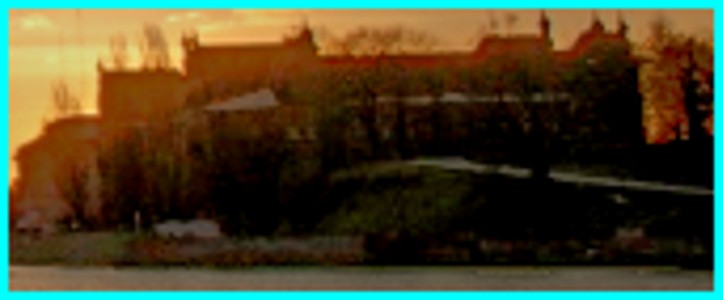}
    \caption{Proposed method with Mertens (Approach 2).
      Entropy: 6.003
      and Naturalness: 0.2975.
      \label{fig:fused_ostrow_variational_mertens}}
  \end{subfigure}
  \begin{subfigure}[t]{0.31\hsize}
    \centering
    \includegraphics[width=\columnwidth]
      {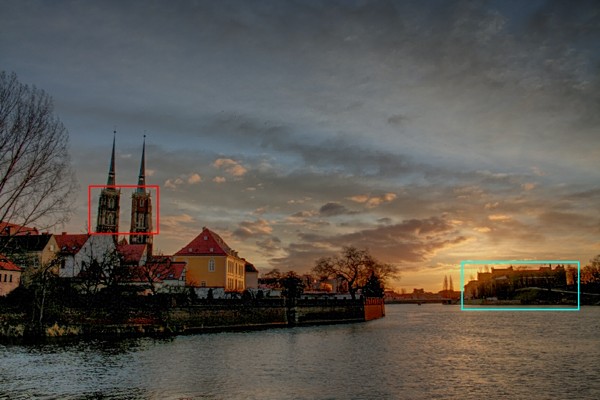}\\
    \includegraphics[width=0.35\columnwidth]
      {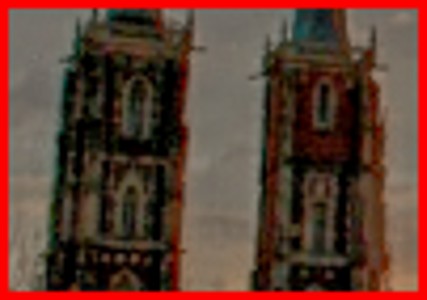}
    \includegraphics[width=0.62\columnwidth]
      {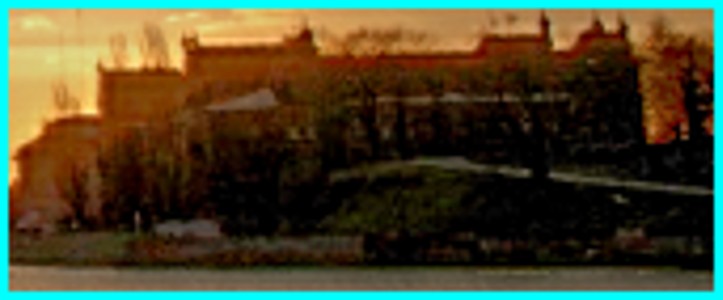}
    \caption{Proposed method with Sakai (Approach 2).
      Entropy: 6.033
      and Naturalness: 0.3274.
      \label{fig:fused_ostrow_variational_sakai}}
  \end{subfigure}
  \begin{subfigure}[t]{0.31\hsize}
    \centering
    \includegraphics[width=\columnwidth]
      {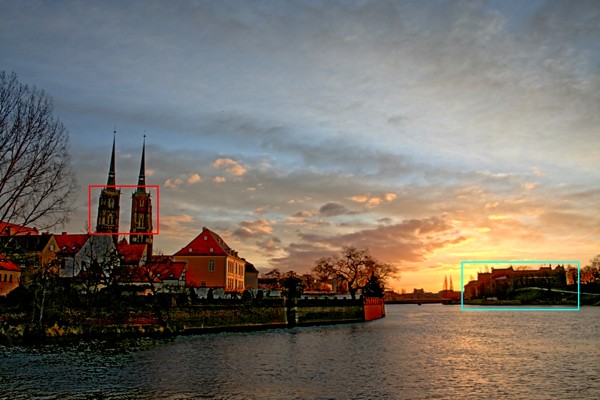}\\
    \includegraphics[width=0.35\columnwidth]
      {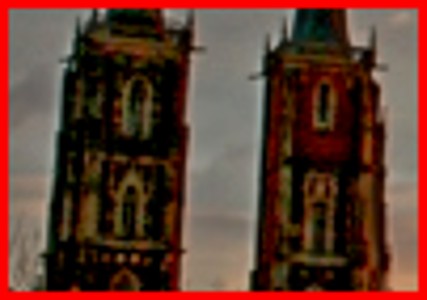}
    \includegraphics[width=0.62\columnwidth]
      {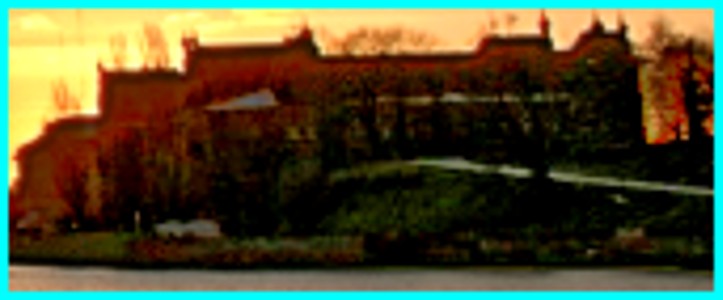}
    \caption{Proposed method with Nejati (Approach 2).
      Entropy: 7.005
      and Naturalness: 0.7568.
      \label{fig:fused_ostrow_variational_nejati}}
  \end{subfigure}
  \caption{Comparison of proposed method with Mertens's, Sakai's, and Nejati's
    methods (``Ostrow Tumski'').
    Zoom-ins of boxed regions are shown in bottom of each fused image.
    Conventional fusion methods without adjustment do not produce clear images
    from unclear multi-exposure images shown in Fig. {\protect\ref{fig:input_ostrow}}
    The proposed SSLA enables us to produce clear images.
    \label{fig:fused_ostrow}}
\end{figure*}

  Figures \ref{fig:fused_window} and \ref{fig:fused_window_li_ma}
  show images fused with/without
  the proposed SSLA from the input multi-exposure ones
  $\{\myvector{x}_n\}$ illustrated in Fig. \ref{fig:input_window}.
  Figures. \ref{fig:fused_window}(\subref{fig:fused_window_org_mertens})
  to (\subref{fig:fused_window_org_nejati}),
  Fig. \ref{fig:fused_window_li_ma}(\subref{fig:fused_window_org_li}),
  and Fig. \ref{fig:fused_window_li_ma}(\subref{fig:fused_window_org_ma})
  show that
  the effects of all of the fusion methods without luminance adjustment
  were not sufficient enough
  to visualize shadow areas when unclear input images were used as inputs.
  In contrast, images fused by using the proposed SSLA clearly represent
  the shadow areas while maintaining the quality of bright areas
  [see Figs. \ref{fig:fused_window}(\subref{fig:fused_window_range_mertens})
  to (\subref{fig:fused_window_variational_nejati})
  and Figs. \ref{fig:fused_window_li_ma}(\subref{fig:fused_window_range_li})
  to (\subref{fig:fused_window_variational_ma})].
  However, combining the proposed SSLA with Li's method
  caused luminance inversion between bright and dark areas.
  Note that the inversion did not occur when Nejati's and Ma's methods,
  which are state-of-the-arts, were combined.
  The results for another image set in Fig. \ref{fig:input_ostrow}
  are also displayed in Fig. \ref{fig:fused_ostrow}.
  Figure \ref{fig:fused_ostrow} shows a similar trend as in Fig. \ref{fig:fused_window}.
  Therefore, the proposed SSLA makes it possible to improve the quality of fused images
  even when input multi-exposure images are unclear.

\begin{figure*}[!t]
  \centering
  \begin{subfigure}[t]{0.25\hsize}
    \centering
    \includegraphics[width=\columnwidth]
      {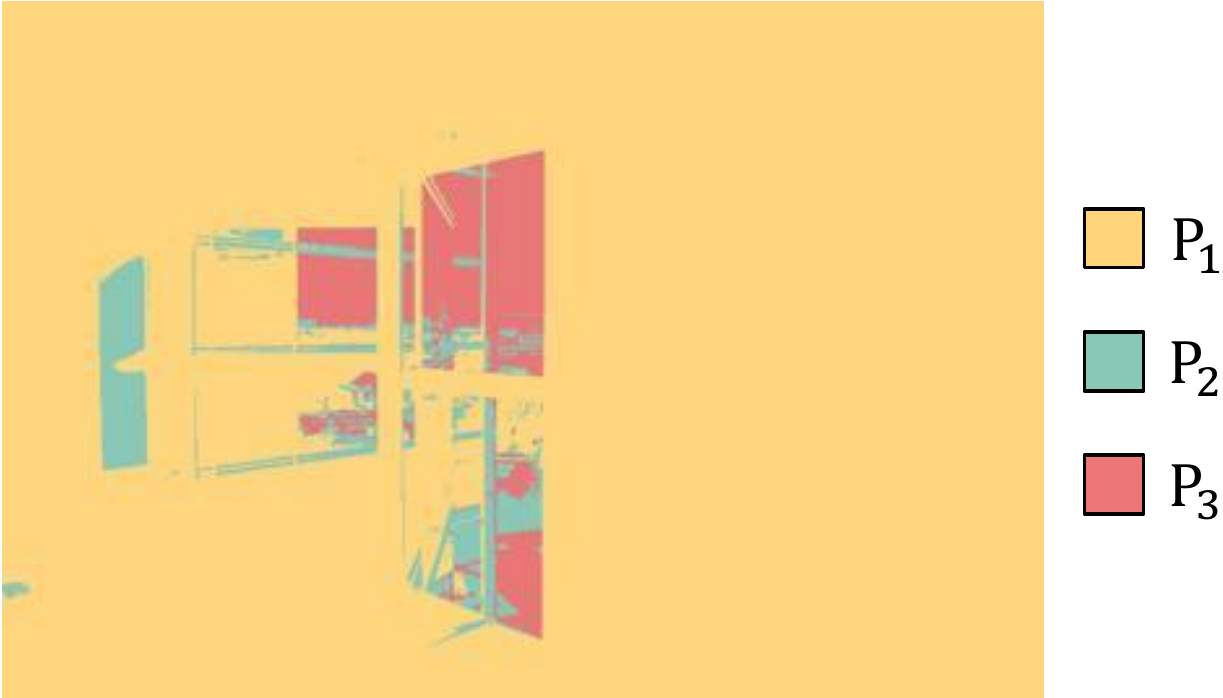}
      \caption{$\{\myset{P}_m\}$ separated by \\
        Approach 1 ($M = 3$)
        \label{fig:window_areas_range}}
  \end{subfigure}
  \hspace{5mm}
  \begin{subfigure}[t]{0.25\hsize}
    \centering
    \includegraphics[width=\columnwidth]
      {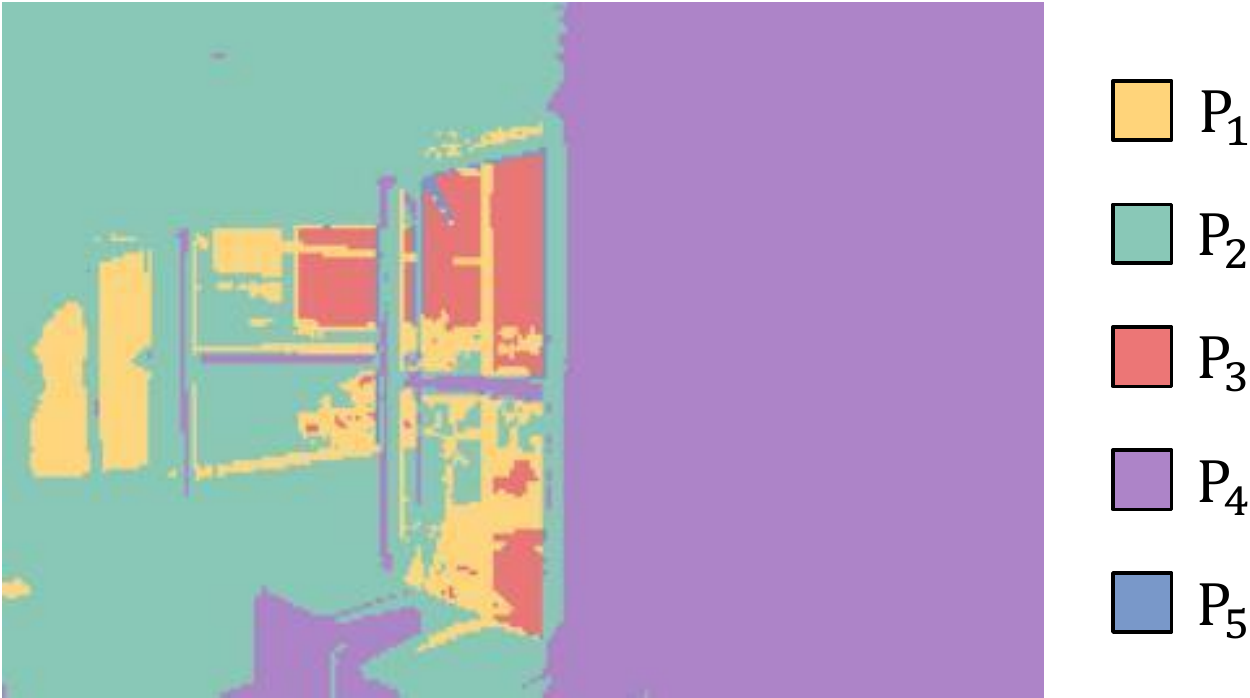}
      \caption{$\{\myset{P}_m\}$ separated by \\
        Approach 2 ($M = 5, K= 10$)
        \label{fig:window_areas_variational}}
  \end{subfigure}\\
  \begin{subfigure}[t]{\hsize}
    \centering
    \includegraphics[width=0.22\columnwidth]
      {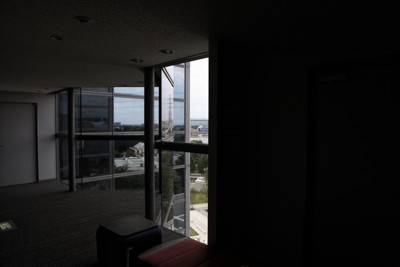}
    \includegraphics[width=0.22\columnwidth]
      {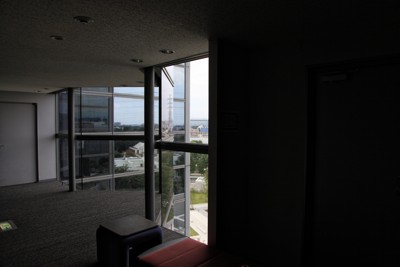}
    \includegraphics[width=0.22\columnwidth]
      {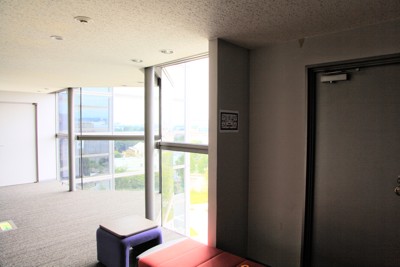}
    \caption{Images $\{\hat{\myvector{x}}_m\}$
      adjusted by using $\{\myset{P}_m\}$
      in Fig. {\protect\ref{fig:adjusted_window}(\subref{fig:window_areas_range})}
      (Approach 1)
      \label{fig:adjusted_window_range}}
  \end{subfigure}\\
  \begin{subfigure}[t]{\hsize}
    \centering
    \includegraphics[width=0.22\columnwidth]
      {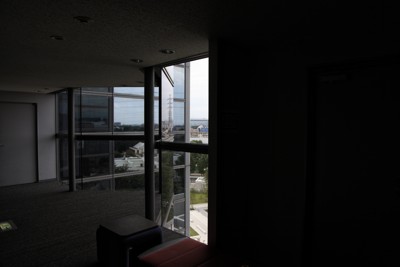}
    \includegraphics[width=0.22\columnwidth]
      {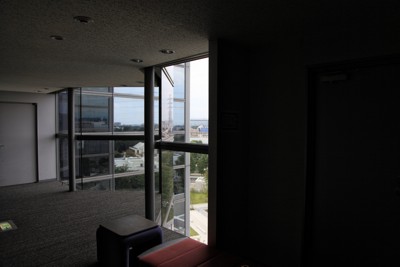}
    \includegraphics[width=0.22\columnwidth]
      {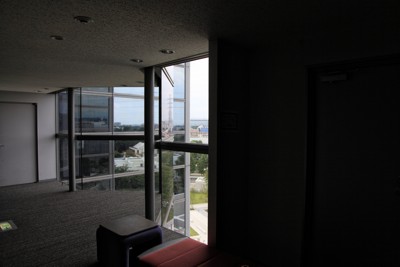}\\
    \includegraphics[width=0.22\columnwidth]
      {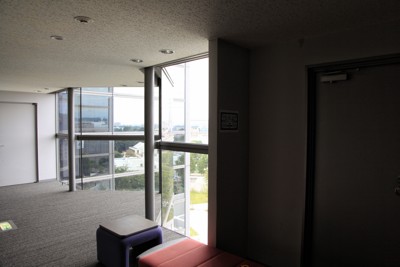}
    \includegraphics[width=0.22\columnwidth]
      {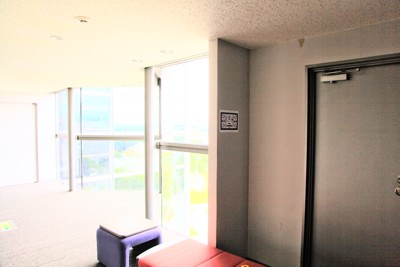}
    \caption{Images $\{\hat{\myvector{x}}_m\}$
      adjusted by using $\{\myset{P}_m\}$
      in Fig. {\protect\ref{fig:adjusted_window}(\subref{fig:window_areas_variational})}
      (Approach 2)
      \label{fig:adjusted_window_variational}}
  \end{subfigure}
  \caption{Adjusted images produced by proposed SSLA (``Window'').
    Approach 1 was not able to separate dark areas on right, e.g., the door,
    and areas on left of middle-brightness, e.g., floor.
    In comparison, Approach 2 was able to separate scene into appropriate areas,
    each having specific luminance range, even though large $K$ was given.
    \label{fig:adjusted_window}}
\end{figure*}

\begin{figure*}[!t]
  \centering
  \begin{subfigure}[t]{0.25\hsize}
    \centering
    \includegraphics[width=\columnwidth]
      {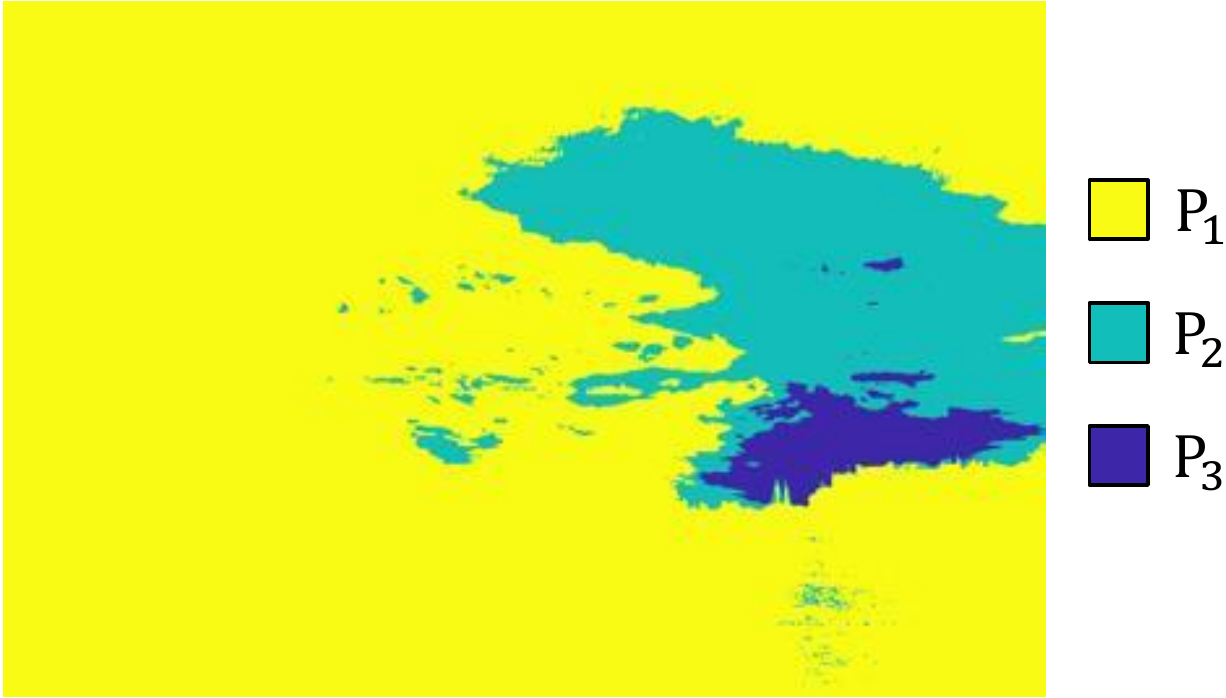}
      \caption{$\{\myset{P}_m\}$ separated by \\
        Approach 1 ($M = 3$)
        \label{fig:ostrow_areas_range}}
  \end{subfigure}
  \hspace{5mm}
  \begin{subfigure}[t]{0.25\hsize}
    \centering
    \includegraphics[width=\columnwidth]
      {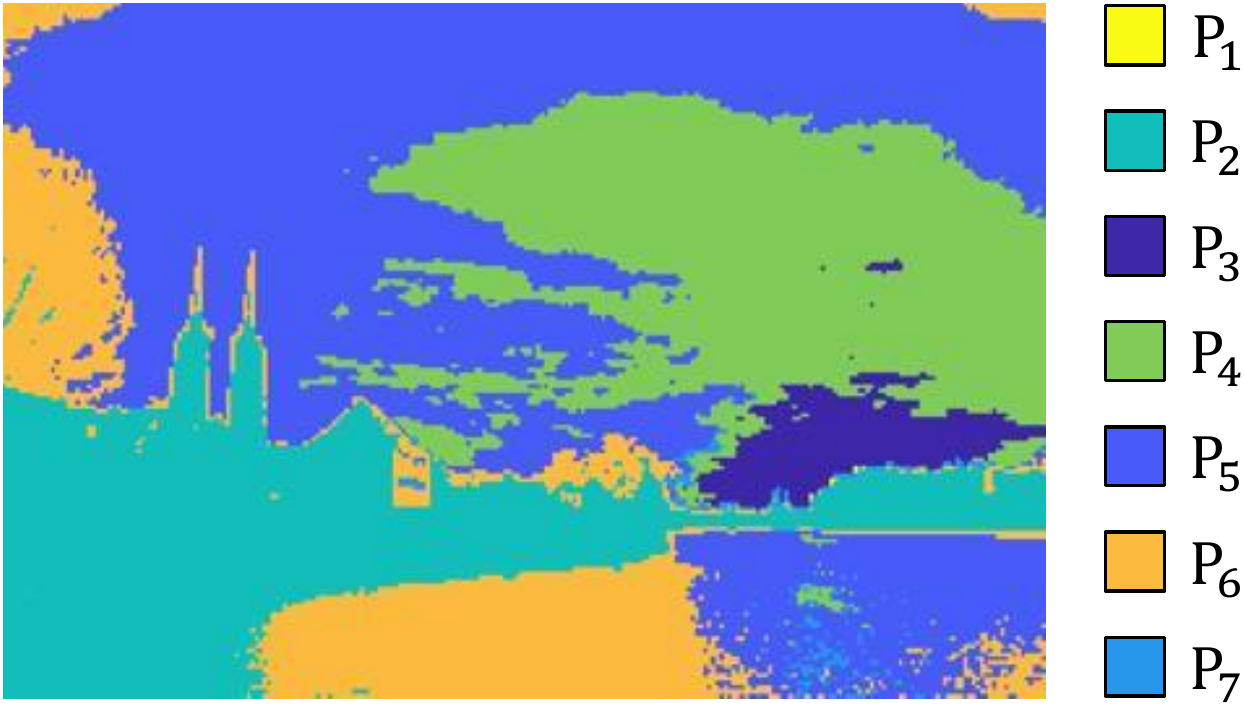}
      \caption{$\{\myset{P}_m\}$ separated by \\
        Approach 2 ($M = 7, K = 10$)
        \label{fig:ostrow_areas_variational}}
  \end{subfigure}\\
  \begin{subfigure}[t]{\hsize}
    \centering
    \includegraphics[width=0.22\columnwidth]
      {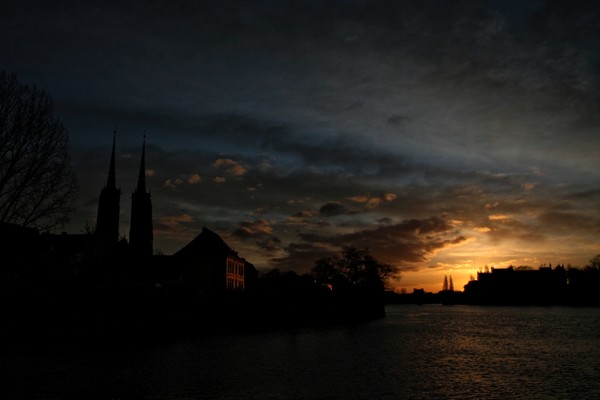}
    \includegraphics[width=0.22\columnwidth]
      {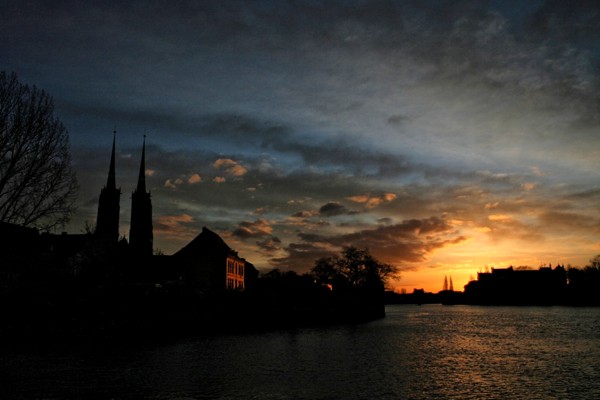}
    \includegraphics[width=0.22\columnwidth]
      {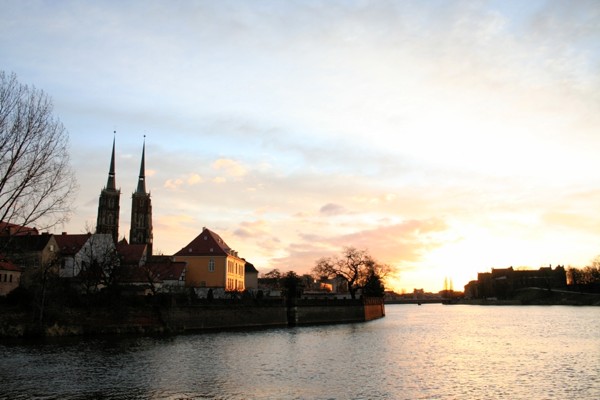}
    \caption{Images $\{\hat{\myvector{x}}_m\}$
      adjusted by using $\{\myset{P}_m\}$
      in Fig. {\protect\ref{fig:adjusted_ostrow}(\subref{fig:ostrow_areas_range})}
      (Approach 1)
      \label{fig:adjusted_ostrow_range}}
  \end{subfigure}\\
  \begin{subfigure}[t]{\hsize}
    \centering
    \includegraphics[width=0.22\columnwidth]
      {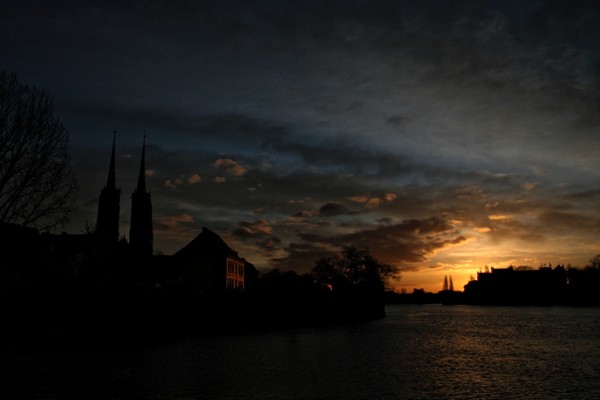}
    \includegraphics[width=0.22\columnwidth]
      {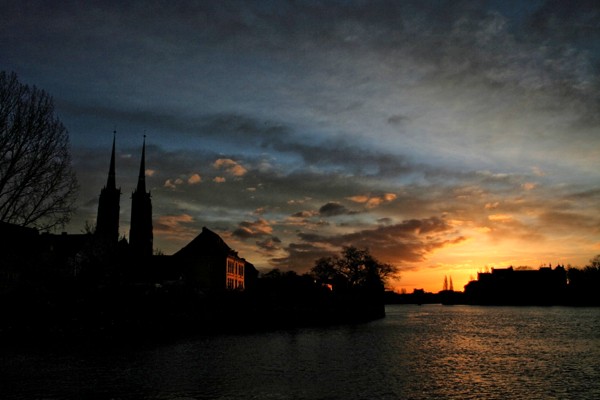}
    \includegraphics[width=0.22\columnwidth]
      {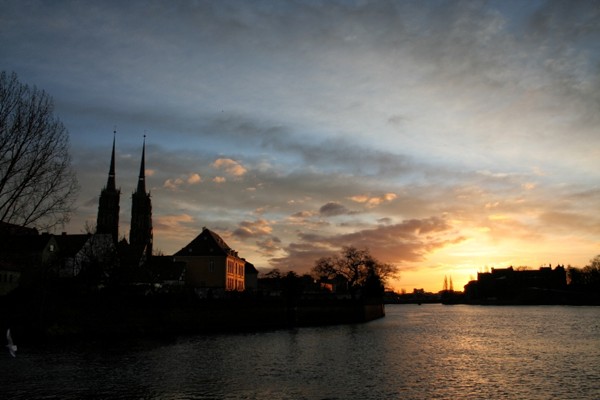}
    \includegraphics[width=0.22\columnwidth]
      {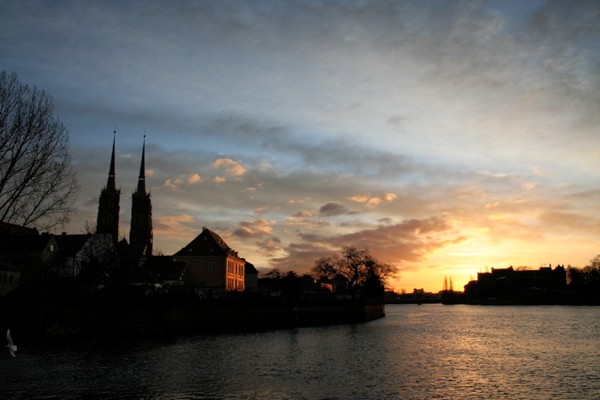}\\
    \includegraphics[width=0.22\columnwidth]
      {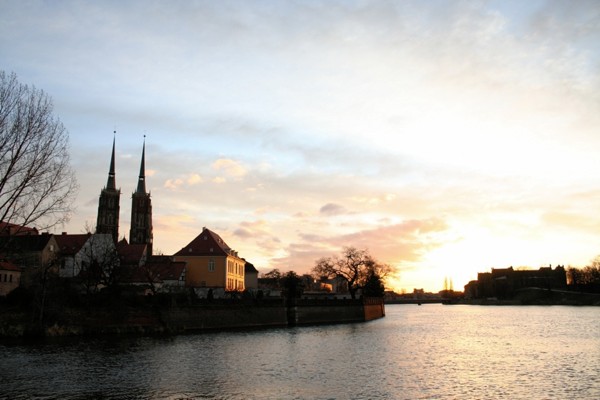}
    \includegraphics[width=0.22\columnwidth]
      {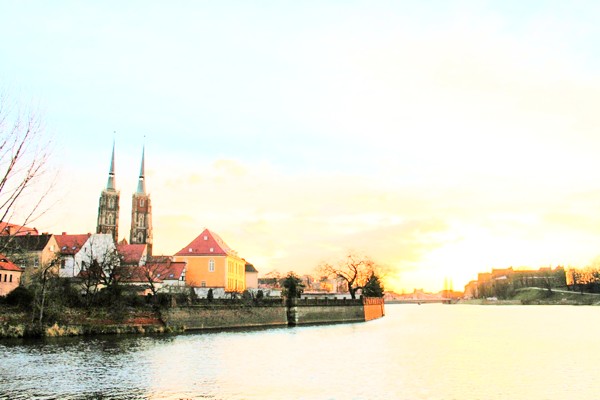}
    \includegraphics[width=0.22\columnwidth]
      {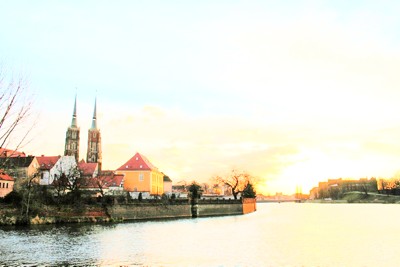}
    \caption{Images $\{\hat{\myvector{x}}_m\}$
      adjusted by using $\{\myset{P}_m\}$
      in Fig. {\protect\ref{fig:adjusted_ostrow}(\subref{fig:ostrow_areas_variational})}
      (Approach 2)
      \label{fig:adjusted_ostrow_variational}}
  \end{subfigure}
  \caption{Adjusted images produced by proposed SSLA (``Ostrow Tumski'').
    Approach 1 was not able to separate sky and waterfront.
    Approach 2 was able to separate scene into appropriate areas
    each having specific luminance range, even though large $K$ was given.
    \label{fig:adjusted_ostrow}}
\end{figure*}

  By comparing Approach 1 with Approach 2,
  Approach 2 generated images representing dark areas more distinctly,
  as shown in Figs. \ref{fig:fused_window} and \ref{fig:fused_ostrow}.
  The difference between the two approaches originates
  in segmented areas $\{\myset{P}_m\}$.
  Figures \ref{fig:adjusted_window} and \ref{fig:adjusted_ostrow} show
  areas $\{\myset{P}_m\}$ separated by Approaches 1 and 2
  and adjusted multi-exposure images based on each $\{\myset{P}_m\}$.
  From Fig. \ref{fig:adjusted_window}(\subref{fig:window_areas_range}),
  it is shown that Approach 1 was not able to separate the dark areas on the right,
  e.g., the door,
  and the areas on the left with middle-brightness, e.g., the floor,
  while it does low computational cost.
  For this reason,
  the brightest image in Fig. \ref{fig:adjusted_window}(\subref{fig:adjusted_window_range})
  does not have sufficient brightness to clearly represent the darkest areas
  in the scene.
  In comparison, Approach 2 was able to separate scenes into appropriate areas,
  each having a specific luminance range, even though a large $K$ was given.
  Figure \ref{fig:adjusted_ostrow} shows almost the same result
  as Fig. \ref{fig:adjusted_window}.
  Hence, it is verified that Approach 2 can separate scenes with higher accuracy
  than Approach 1, while Approach 1 can be performed with closed-form calculation.

\begin{figure}[!t]
  \centering
  \begin{subfigure}[t]{0.9\hsize}
    \centering
    \includegraphics[width=\columnwidth]
      {./figs/window_evaluative_nejati_variational_fused.jpg}\\
    \includegraphics[width=0.34\columnwidth]
      {./figs/window_evaluative_nejati_variational_patch1.jpg}
    \includegraphics[width=0.63\columnwidth]
      {./figs/window_evaluative_nejati_variational_patch2.jpg}
    \caption{Approach 2 with contrast enhancement.\\
      Entropy: 6.848
      and Naturalness: 0.7031.
      \label{fig:fused_window_variational_nejati_ce}}
  \end{subfigure}\\
  \begin{subfigure}[t]{0.9\hsize}
    \centering
    \includegraphics[width=\columnwidth]
      {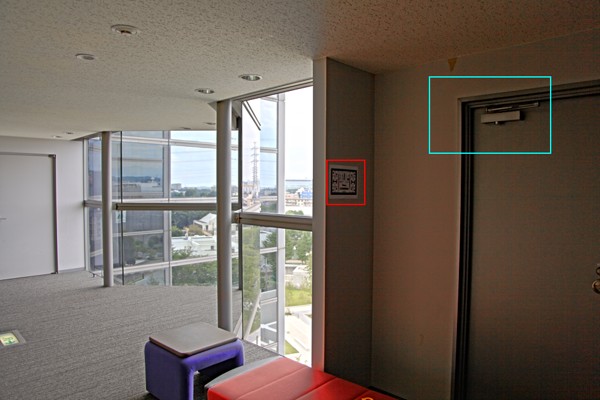}\\
    \includegraphics[width=0.34\columnwidth]
      {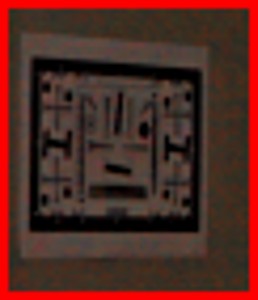}
    \includegraphics[width=0.63\columnwidth]
      {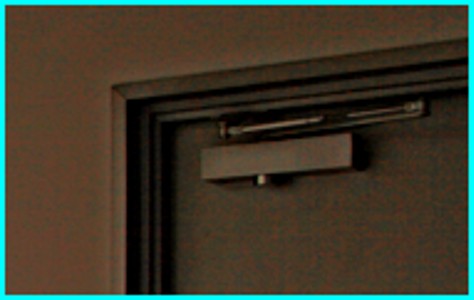}
    \caption{Approach 2 without contrast ehnancement.\\
      Entropy: 6.810
      and Naturalness: 0.5943.
      \label{fig:fused_window_variational_nejati_wo_ce}}
  \end{subfigure}\\
  \caption{Effect of local contrast enhancement in proposed method (``Window'').
    Nejati's fusion method \cite{nejati2017fast} and proposed Approach 2 were utilized.
    Zoom-ins of boxed regions are shown in bottom of each fused image.
    Conventional fusion methods without adjustment do not produce clear images
    from unclear multi-exposure images shown in Fig. {\protect\ref{fig:input_window}}.
    Local contrast enhancement can make image structure to be clear,
    but it often causes noise and ringing to be boosted.
    \label{fig:fused_window_ce}}
\end{figure}

  Figures \ref{fig:adjusted_window} and \ref{fig:adjusted_ostrow} also illustrate
  that both Approach 1 and 2 maintained the structure of images even though
  the two approaches did not pay attention to the structure.
  This result represents that the segmentation method in the proposed SSLA
  does not require the information of image structure such as edges.
  Figure \ref{fig:fused_window_ce} shows the effect of
  local contrast enhancement in the proposed method.
  This enhancement can reduce blurring and make the structure in images clear,
  but it often causes noise and ringing to be boosted.
  In this case, the proposed method without local contrast enhancement is effective
  to suppress such noise boosting.

\begin{figure*}[!t]
  \centering
  \begin{subfigure}[t]{0.24\hsize}
    \centering
    \includegraphics[width=\columnwidth]
      {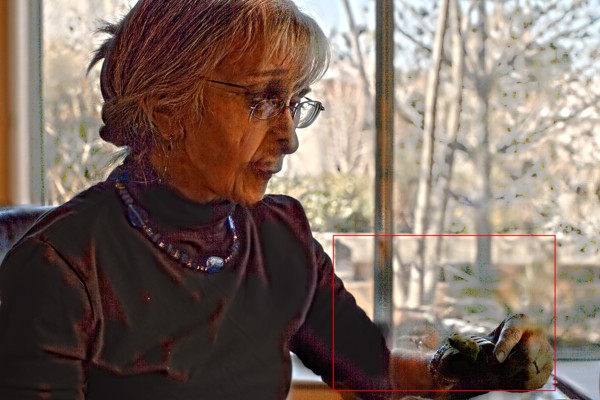}\\
    \includegraphics[width=\columnwidth]
      {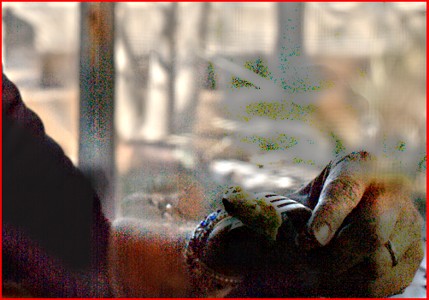}\\
    \caption{Ma {\protect\cite{ma2017robust}}.
      Entropy: 6.917\\
      and Naturalness: 0.8262.
      \label{fig:fused_lady_org_ma}}
  \end{subfigure}
  \begin{subfigure}[t]{0.24\hsize}
    \centering
    \includegraphics[width=\columnwidth]
      {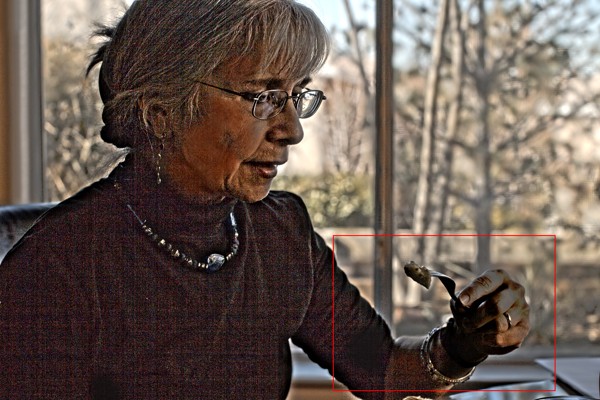}\\
    \includegraphics[width=\columnwidth]
      {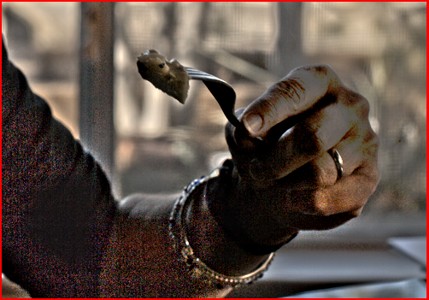}\\
    \caption{Proposed method with Ma (Approach 1).
      Entropy: 6.653\\
      and Naturalness: 0.6731.
      \label{fig:fused_lady_range_ma}}
  \end{subfigure}
  \begin{subfigure}[t]{0.24\hsize}
    \centering
    \includegraphics[width=\columnwidth]
      {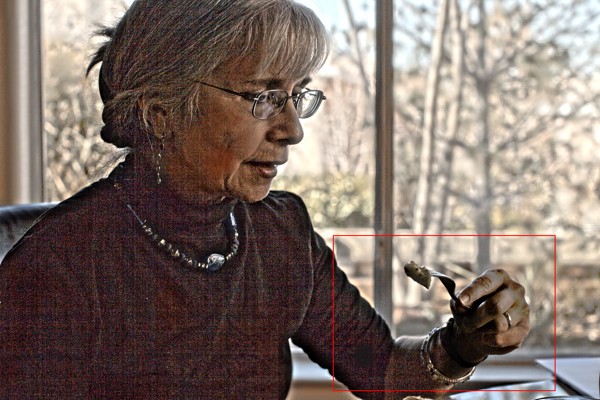}\\
    \includegraphics[width=\columnwidth]
      {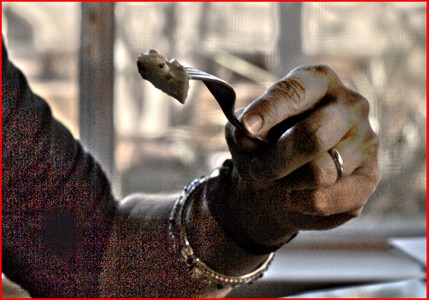}\\
    \caption{Proposed method with Ma (Approach 2).
      Entropy: 7.157\\
      and Naturalness: 0.9701.
      \label{fig:fused_lady_variational_ma}}
  \end{subfigure}
  \begin{subfigure}[t]{0.24\hsize}
    \centering
    \includegraphics[width=\columnwidth]
      {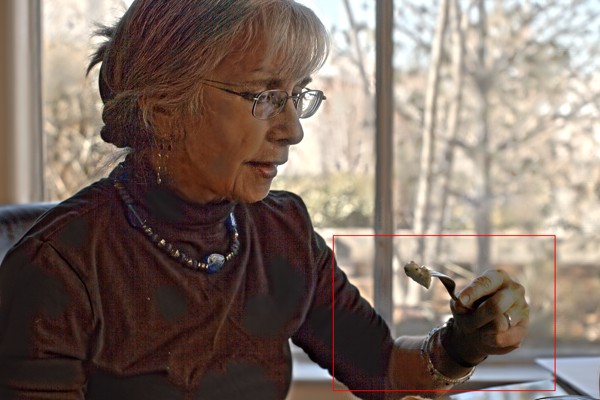}\\
    \includegraphics[width=\columnwidth]
      {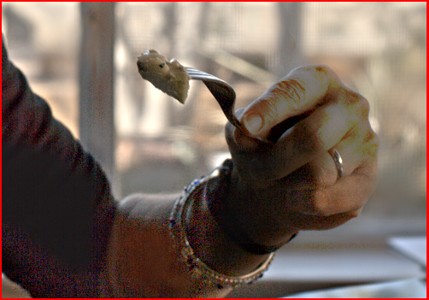}\\
    \caption{Proposed method with Ma (Approach 2
      without local contrast enhancement).\\
      Entropy: 7.000\\
      and Naturalness: 0.7076.
      \label{fig:fused_lady_variational_wo_ce_ma}}
  \end{subfigure}
  \caption{Comparison of proposed method with Ma's fusion method (``Lady eating'').
    Zoom-ins of boxed region are shown in bottom of each fused image.
    The proposed SSLS is effective for dynamic scenes.
    \label{fig:fused_lady}}
\end{figure*}

  Fused images from input multi-exposure ones containing moving objects
  are shown in Fig. \ref{fig:fused_lady}.
  Even when moving objects are included in input images,
  the proposed method can still provide high-quality fused images without ghost-like artifacts
  by using a robust method to deal with dynamic scenes as fusion method $\mathscr{F}$.
  In contrast, the image generated without the proposed method
  contained a few ghost-like artifacts.
  This result indicates that the proposed method is able
  not only to clarify fused images,
  but also to remove ghost-like artifacts.
\subsection{Quantitative evaluation}
  To quantitatively evaluate the quality of fused images,
  objective quality metrics are needed.
  Typical quality metrics such as the peak signal to noise ratio (PSNR)
  and structural similarity index (SSIM) are not suitable for this purpose
  because they use a target image with the highest quality as a reference.
  Recently, some full-reference metrics for multi-exposure image fusion
  have been proposed \cite{ma2015perceptual, rahman2017evaluating}.
  However, because high-quality multi-exposure images are needed for these metrics,
  they are also not suitable for evaluating the proposed method
  when the multi-exposure ones are unavailable.
  We therefore used the metric \cite{ma2015perceptual}, referred to as MEF-SSIM,
  only when high-quality multi-exposure ones are available.
  In other cases, the discrete entropy of luminance values and
  tone mapped image quality index (TMQI) \cite{yeganeh2013objective}
  were used as quality metrics.

  MEF-SSIM is based on a multi-scale SSIM framework and a patch consistency measure.
  It keeps a good balance between local structure preservation
  and global luminance consistency.
  Discrete entropy represents the amount of information in an image,
  where the entropy is calculated
  by using the luminance values of a fused image $\myvector{y}$.
  Discrete entropy of luminance also shows the global contrast of an image.
  TMQI represents the quality of an image tone-mapped from an HDR image;
  the metric incorporates structural fidelity and statistical naturalness.
  Statistical naturalness \cite{yeganeh2013objective} is defined as
  \begin{equation}
    S = \frac{1}{K} \gaussian{\overline{l}}{115.94}{27.99^2}
        \betapdf{\overline{\sigma_l}/64.29}{4.4}{10.1},
  \end{equation}
  where $K$ is a normalization factor,
  $\betapdf{\cdot}{\alpha}{\beta}$ is a beta probability density function
  with parameters $\alpha$ and $\beta$,
  $\overline{l}$ is the average luminance of an image on a scale $[0, 255]$,
  and $\overline{\sigma_l}$ is the average of local standard deviation of luminance
  on the scale.
  The Gaussian and beta distributions reflect naturalness of the global brightness
  and local contrast, respectively.
  Statistical naturalness is calculated without any reference images,
  although structural fidelity needs an HDR image as a reference.
  Since the processes of tone mapping and photographing are similar,
  TMQI is also useful for evaluating photographs.
  In this simulation, MEF-SSIM and TMQI were used only for evaluation
  with the sets of tone-mapped images,
  while discrete entropy and statistical naturalness were used for evaluation
  with all of the 570 sets of images.

  Tables \ref{tab:entropy_photograph} and \ref{tab:naturalness_photograph}
  summarize scores for the 12 sets of photographs of static scenes in terms of
  discrete entropy and statistical naturalness.
  For each score (discrete entropy $\in [0, 8]$
  and statistical naturalness $\in [0, 1]$),
  a larger value means higher quality.
  Table \ref{tab:entropy_photograph} shows that
  the proposed SSLA provided high entropy scores for all 12 image sets,
  although the typical fusion methods provided low entropy scores for ``Corridor 1''
  and ``Lobby.''
  In addition, the proposed SSLA provided higher average scores than
  the typical fusion methods.
  Table \ref{tab:naturalness_photograph} denotes that
  the proposed SSLA enables us to produce high-quality images in terms of
  statistical naturalness in almost all cases, compared with the typical methods.
\begin{table*}[!t]
  \centering
  \caption{Discrete entropy scores for static scenes.
  Boldface indicates higher score.
  ``w/o'' means that images $\myvector{y}$ were produced without luminance adjustment.
  ``Prop. 1'' and ``Prop. 2'' indicate proposed method using Approach 1
  and proposed method using Approach 2, respectively.}
  {
  \footnotesize
  \tabcolsep = 1mm
  \begin{tabular}{l|ccc|ccc|ccc|ccc|ccc}\hline\hline
    \multirow{2}{*}{Scene} & \multicolumn{3}{|c|}{Mertens \cite{mertens2009exposure}}
      & \multicolumn{3}{|c|}{Sakai \cite{sakai2015hybrid}}
      & \multicolumn{3}{|c|}{Nejati \cite{nejati2017fast}}
      & \multicolumn{3}{|c|}{Li \cite{li2013image}}
      & \multicolumn{3}{|c}{Ma \cite{ma2017robust}}\\ \cline{2-16}
      & w/o & Prop. 1 & Prop. 2 & w/o & Prop. 1 & Prop. 2 & w/o & Prop. 1 & Prop. 2
    & w/o & Prop. 1 & Prop. 2 & w/o & Prop. 1 & Prop. 2 \\ \hline
    Arno & \textbf{5.987} & 5.931 & 5.833 & \textbf{5.992} & 5.935 & 5.844
     & \textbf{7.016} & \textbf{7.016} & 6.793 & \textbf{6.312} & 6.095 & 6.072
     & 7.054 & \textbf{7.100} & 6.813 \\
    Cave & 4.938 & 5.920 & \textbf{6.277} & 4.946 & 5.942 & \textbf{6.290}
     & 6.132 & 6.839 & \textbf{7.044} & 6.115 & 6.840 & \textbf{6.999}
     & 6.082 & 6.782 & \textbf{7.043} \\
    Chinese garden & 6.142 & 6.185 & \textbf{6.241} & 6.155 & 6.205 & \textbf{6.258}
     & 6.946 & \textbf{7.215} & 6.844 & 7.244 & \textbf{7.271} & 7.229
     & 6.998 & \textbf{7.143} & 6.955 \\
    Corridor 1 & 2.858 & \textbf{5.840} & 5.738 & 2.856 & \textbf{5.859} & 5.752
     & 2.752 & \textbf{7.134} & 6.976 & 3.350 & 7.242 & \textbf{7.262}
     & 2.773 & \textbf{7.088} & 7.016 \\
    Corridor 2 & 5.706 & \textbf{5.922} & 5.613 & 5.714 & \textbf{5.933} & 5.618
     & 6.543 & \textbf{6.925} & 6.625 & \textbf{6.682} & 5.548 & 5.670
     & 6.533 & \textbf{6.930} & 6.586 \\
    Estate rsa & 6.577 & \textbf{6.604} & 6.585 & 6.560 & \textbf{6.584} & 6.573
     & 6.785 & 6.917 & \textbf{7.014} & 7.001 & \textbf{7.015} & 6.914
     & 6.802 & 6.921 & \textbf{7.014} \\
    Kluki & \textbf{7.120} & 6.982 & 6.942 & \textbf{7.136} & 7.003 & 6.964
     & \textbf{7.489} & 7.178 & 7.151 & \textbf{7.669} & 7.540 & 7.548
     & \textbf{7.535} & 7.266 & 7.207 \\
    Laurenziana & \textbf{6.672} & 6.478 & 6.599 & \textbf{6.678} & 6.490 & 6.609
     & \textbf{7.153} & 6.956 & 6.898 & \textbf{7.585} & 7.239 & 7.417
     & \textbf{7.267} & 7.091 & 7.008 \\
    Lobby & 4.172 & 5.648 & \textbf{5.737} & 4.177 & 5.672 & \textbf{5.743}
     & 4.929 & \textbf{7.118} & 6.836 & 5.274 & 6.773 & \textbf{7.177}
     & 4.893 & \textbf{7.060} & 6.783 \\
    Mountains & \textbf{6.920} & 6.480 & 6.571 & \textbf{6.912} & 6.492 & 6.586
     & 6.823 & 6.790 & \textbf{6.846} & \textbf{6.491} & 6.233 & 6.339
     & 6.796 & 6.781 & \textbf{6.838} \\
    Ostrow tumski & 5.861 & 5.607 & \textbf{6.003} & 5.874 & 5.629 & \textbf{6.033}
     & 6.812 & 6.475 & \textbf{7.005} & 6.607 & 6.253 & \textbf{6.794}
     & 6.857 & 6.622 & \textbf{6.958} \\
    Window & 5.047 & 5.715 & \textbf{6.060} & 5.052 & 5.735 & \textbf{6.072}
     & 5.407 & 6.750 & \textbf{6.848} & 5.547 & 6.617 & \textbf{7.099}
     & 5.402 & 6.752 & \textbf{6.868} \\ \hdashline
    Average & 5.667 & 6.109 & \textbf{6.183} & 5.671 & 6.123 & \textbf{6.195}
     & 6.232 & \textbf{6.943} & 6.907 & 6.323 & 6.722 & \textbf{6.877}
     & 6.249 & \textbf{6.961} & 6.924 \\ \hline
\end{tabular}
  }
  \label{tab:entropy_photograph}
\end{table*}
\begin{table*}[!t]
  \centering
  \caption{Statistical naturalness scores for static scenes.
  Boldface indicates higher score.
  ``w/o'' means that images $\myvector{y}$ were produced without luminance adjustment.
  ``Prop. 1'' and ``Prop. 2'' indicate proposed method using Approach 1
  and proposed method using Approach 2, respectively.}
  {
  \footnotesize
  \tabcolsep = 1mm
  \begin{tabular}{l|ccc|ccc|ccc|ccc|ccc}\hline\hline
    \multirow{2}{*}{Scene} & \multicolumn{3}{|c|}{Mertens \cite{mertens2009exposure}}
      & \multicolumn{3}{|c|}{Sakai \cite{sakai2015hybrid}}
      & \multicolumn{3}{|c|}{Nejati \cite{nejati2017fast}}
      & \multicolumn{3}{|c|}{Li \cite{li2013image}}
      & \multicolumn{3}{|c}{Ma \cite{ma2017robust}}\\ \cline{2-16}
      & w/o & Prop. 1 & Prop. 2 & w/o & Prop. 1 & Prop. 2 & w/o & Prop. 1 & Prop. 2
     & w/o & Prop. 1 & Prop. 2 & w/o & Prop. 1 & Prop. 2 \\ \hline
    Arno & 0.1289 & \textbf{0.1382} & 0.1233 & 0.1506 & \textbf{0.1633} & 0.1473
     & 0.5164 & 0.5372 & \textbf{0.5442} & \textbf{0.4079} & 0.2852 & 0.3543
     & 0.5076 & \textbf{0.6845} & 0.5851 \\
    Cave & 0.0758 & 0.2458 & \textbf{0.2808} & 0.0758 & 0.2236 & \textbf{0.2512}
     & \textbf{0.4223} & 0.3413 & 0.0628 & \textbf{0.2665} & 0.2230 & 0.0347
     & \textbf{0.4122} & 0.3261 & 0.0758 \\
    Chinese garden & 0.420 & 0.4353 & \textbf{0.4612} & 0.4059 & 0.4222 & \textbf{0.4441}
     & 0.4609 & \textbf{0.4843} & 0.4152 & 0.4801 & \textbf{0.5316} & 0.4477
     & \textbf{0.4771} & 0.4468 & 0.3863 \\
    Corridor 1 & 0.0004 & \textbf{0.1454} & 0.1374 & 0.0004 & \textbf{0.1998} & 0.1713
     & 0.0006 & \textbf{0.9111} & 0.8792 & 0.0012 & 0.9017 & \textbf{0.9129}
     & 0.0006 & \textbf{0.8908} & 0.8825 \\
    Corridor 2 & 0.0628 & \textbf{0.1025} & 0.0744 & 0.0763 & \textbf{0.1338} & 0.0911
     & 0.3368 & \textbf{0.5052} & 0.4459 & \textbf{0.3807} & 0.0957 & 0.1293
     & 0.3212 & \textbf{0.5209} & 0.4632 \\
    Estate rsa & 0.7919 & \textbf{0.8047} & 0.7992 & 0.7624 & 0.7755 & \textbf{0.7840}
     & 0.8884 & 0.9504 & \textbf{0.9872} & 0.9773 & 0.9894 & \textbf{0.9941}
     & 0.9006 & 0.9536 & \textbf{0.9903} \\
    Kluki & \textbf{0.9665} & 0.9580 & 0.9105 & \textbf{0.9596} & 0.9432 & 0.8889
     & 0.9316 & \textbf{0.9844} & 0.9749 & 0.5148 & \textbf{0.7616} & 0.7226
     & 0.8659 & 0.9849 & \textbf{0.9940}  \\
    Laurenziana & \textbf{0.8046} & 0.6541 & 0.7387 & \textbf{0.8058} & 0.6595 & 0.7366
     & 0.9330 & \textbf{0.9722} & 0.9558 & 0.6125 & \textbf{0.9408} & 0.8303
     & 0.8902 & \textbf{0.9662} & 0.9319 \\
    Lobby & 0.0167 & 0.1947 & \textbf{0.2093} & 0.0190 & \textbf{0.2307} & 0.2274
     & 0.1019 & \textbf{0.9260} & 0.7637 & 0.1350 & 0.6566 & \textbf{0.6936}
     & 0.1025 & \textbf{0.8888} & 0.7285 \\
    Mountains & 0.3953 & 0.5582 & \textbf{0.6231} & 0.3562 & 0.5795 & \textbf{0.6353}
     & 0.4702 & 0.7383 & \textbf{0.7634} & 0.4852 & 0.5860 & \textbf{0.6898}
     & 0.4373 & 0.7459 & \textbf{0.8171} \\
    Ostrow tumski & 0.1445 & 0.1347 & \textbf{0.2975} & 0.1580 & 0.1516 & \textbf{0.3274}
     & 0.5090 & 0.6194 & \textbf{0.7568} & 0.4874 & 0.3090 & \textbf{0.8185}
     & 0.5145 & 0.7585 & \textbf{0.8162} \\
    Window & 0.0569 & 0.0912 & \textbf{0.2135} & 0.0615 & 0.1142 & \textbf{0.2650}
     & 0.1617 & 0.6187 & \textbf{0.7031} & 0.2145 & 0.5070 & \textbf{0.8770}
     & 0.1655 & 0.6413 & \textbf{0.7475} \\ \hdashline
    Average & 0.3220 & 0.3719 & \textbf{0.4057} & 0.3193 & 0.3831 & \textbf{0.4141}
     & 0.4777 & \textbf{0.7157} & 0.6877 & 0.4136 & 0.5657 & \textbf{0.6254}
     & 0.4663 & \textbf{0.734}  & 0.7015 \\ \hline
\end{tabular}
  }
  \label{tab:naturalness_photograph}
\end{table*}

  Tables \ref{tab:entropy_dynamic} and \ref{tab:naturalness_dynamic} denote scores
  for the eight sets of photographs of dynamic scenes in terms of discrete entropy
  and statistical naturalness.
  Here, Ma's method was used as fusion method $\mathscr{F}$.
  Since the sets contained enough number of multi-exposure images,
  the scores did not have a large difference between with/without the proposed method.
\begin{table}[!t]
  \centering
  \caption{Discrete entropy scores for dynamic scenes.
  Boldface indicates higher score.
  ``w/o'' means that images $\myvector{y}$ were produced without luminance adjustment.
  ``Prop. 1'' and ``Prop. 2'' indicate proposed method using Approach 1
  and proposed method using Approach 2, respectively.}
  \begin{tabular}{l|cccc}\hline\hline
Scene                & w/o            & Prop. 1        & Prop. 2        & Prop. 2 w/o CE \\ \hline
Baby at window       & 7.032          & 6.351          & \textbf{7.160} & 6.770          \\
Baby on grass        & 7.072          & 6.711          & \textbf{7.271} & 6.966          \\
Christmas rider      & \textbf{7.098} & 6.366          & 7.048          & 6.945          \\
Feeding time         & 6.498          & \textbf{6.853} & 6.643          & 6.621          \\
High chair           & \textbf{7.272} & 6.861          & 7.163          & 7.149          \\
Lady eating          & 6.917          & 6.653          & \textbf{7.157} & 7.000          \\
Piano man            & 6.633          & 6.632          & \textbf{6.791} & 6.752          \\
Santas little helper & 7.186          & 6.429          & 7.246          & \textbf{7.285} \\ \hdashline
Average              & 6.964          & 6.607          & \textbf{7.060} & 6.936         \\ \hline
\end{tabular}
\label{tab:entropy_dynamic}
\end{table}
\begin{table}[!t]
  \centering
  \caption{Statistical naturalness scores for dynamic scenes.
  Boldface indicates higher score.
  ``w/o'' means that images $\myvector{y}$ were produced without luminance adjustment.
  ``Prop. 1'' and ``Prop. 2'' indicate proposed method using Approach 1
  and proposed method using Approach 2, respectively.}
  \begin{tabular}{l|cccc}\hline\hline
Scene                & w/o             & Prop. 1         & Prop. 2         & Prop. 2 w/o CE  \\ \hline
Baby at window       & \textbf{0.9912} & 0.5344          & 0.9830          & 0.9438          \\
Baby on grass        & \textbf{0.9779} & 0.6881          & 0.8532          & 0.8629          \\
Christmas rider      & 0.8797          & 0.5972          & 0.7192          & \textbf{0.9593} \\
Feeding time         & 0.1699          & \textbf{0.7162} & 0.6393          & 0.2859          \\
High chair           & \textbf{0.9549} & 0.6721          & 0.7856          & 0.8372          \\
Lady eating          & 0.8262          & 0.6731          & \textbf{0.9701} & 0.7076          \\
Piano man            & 0.6369          & 0.6016          & 0.2942          & \textbf{0.7637} \\
Santas little helper & 0.6302          & 0.3866          & 0.3271          & \textbf{0.9089} \\ \hdashline
Average              & 0.7584          & 0.6087          & 0.6965          & \textbf{0.7837} \\ \hline
\end{tabular}
\label{tab:naturalness_dynamic}
\end{table}
  
\begin{figure*}[!t]
  \centering
  \begin{subfigure}[t]{0.24\hsize}
    \centering
    \includegraphics[width=\columnwidth]{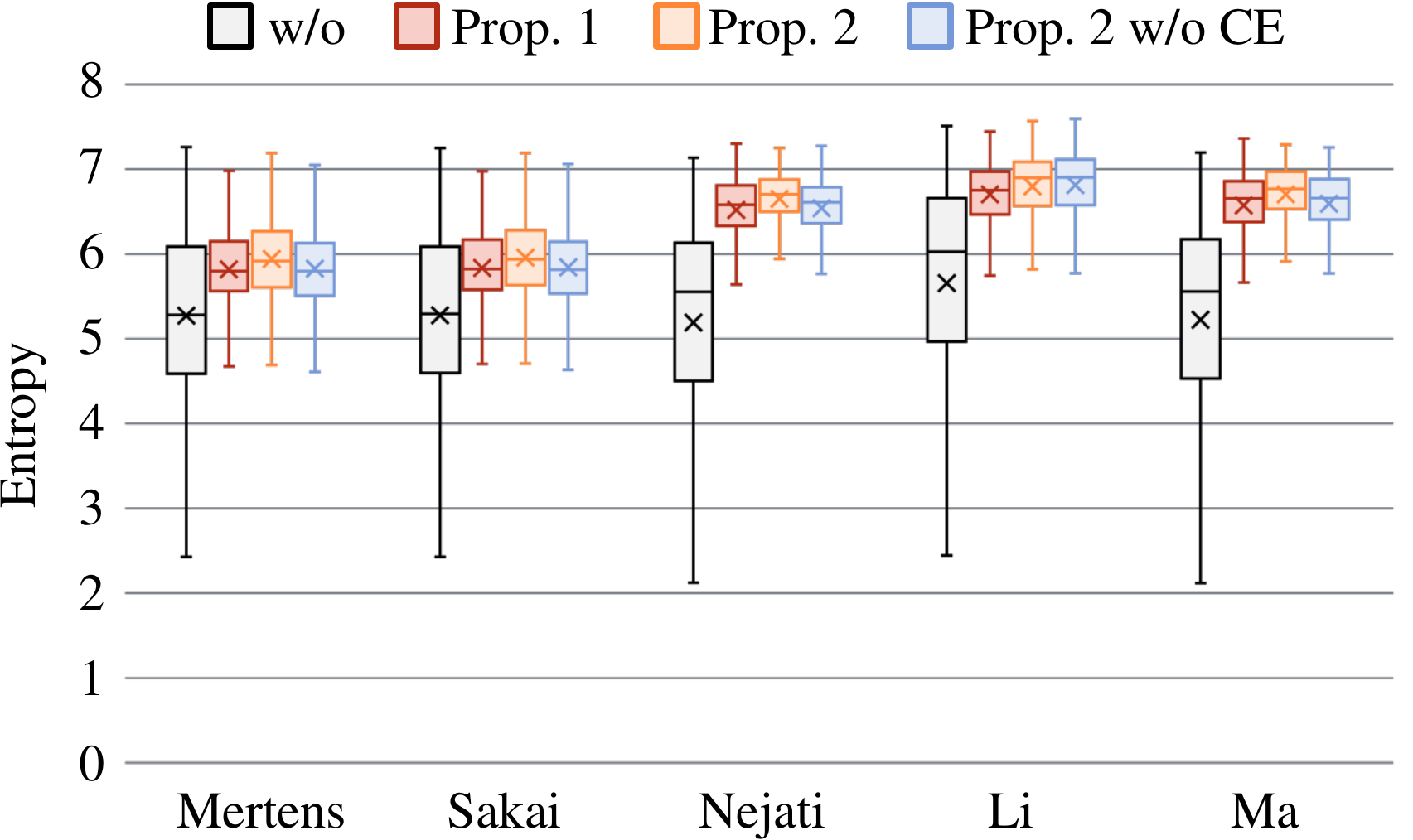}
    \caption{Discrete entropy
      \label{fig:entropy_random}}
  \end{subfigure}
  \begin{subfigure}[t]{0.24\hsize}
    \centering
    \includegraphics[width=\columnwidth]{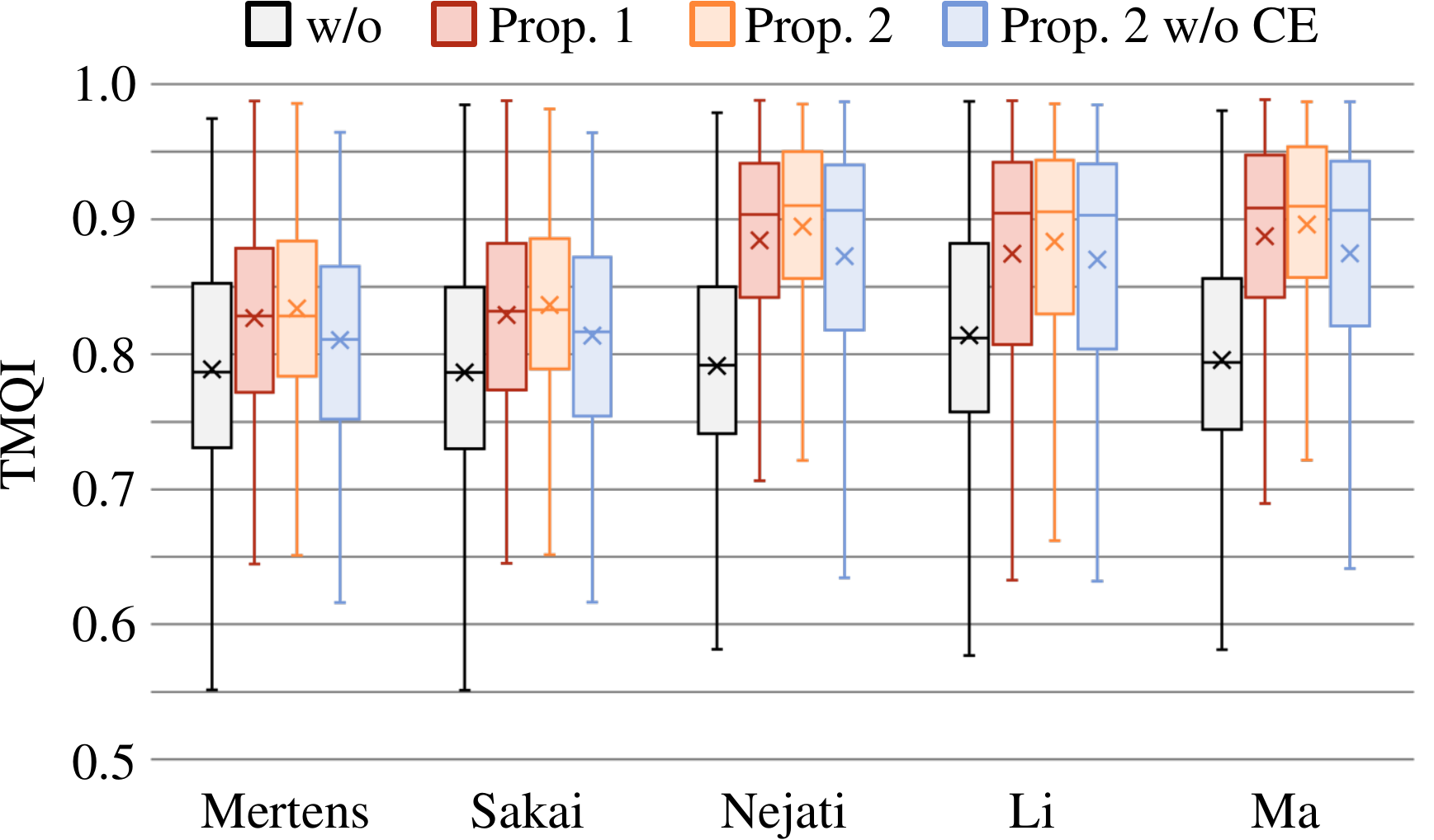}
    \caption{TMQI
      \label{fig:tmqi_random}}
  \end{subfigure}
  \begin{subfigure}[t]{0.24\hsize}
    \centering
    \includegraphics[width=\columnwidth]{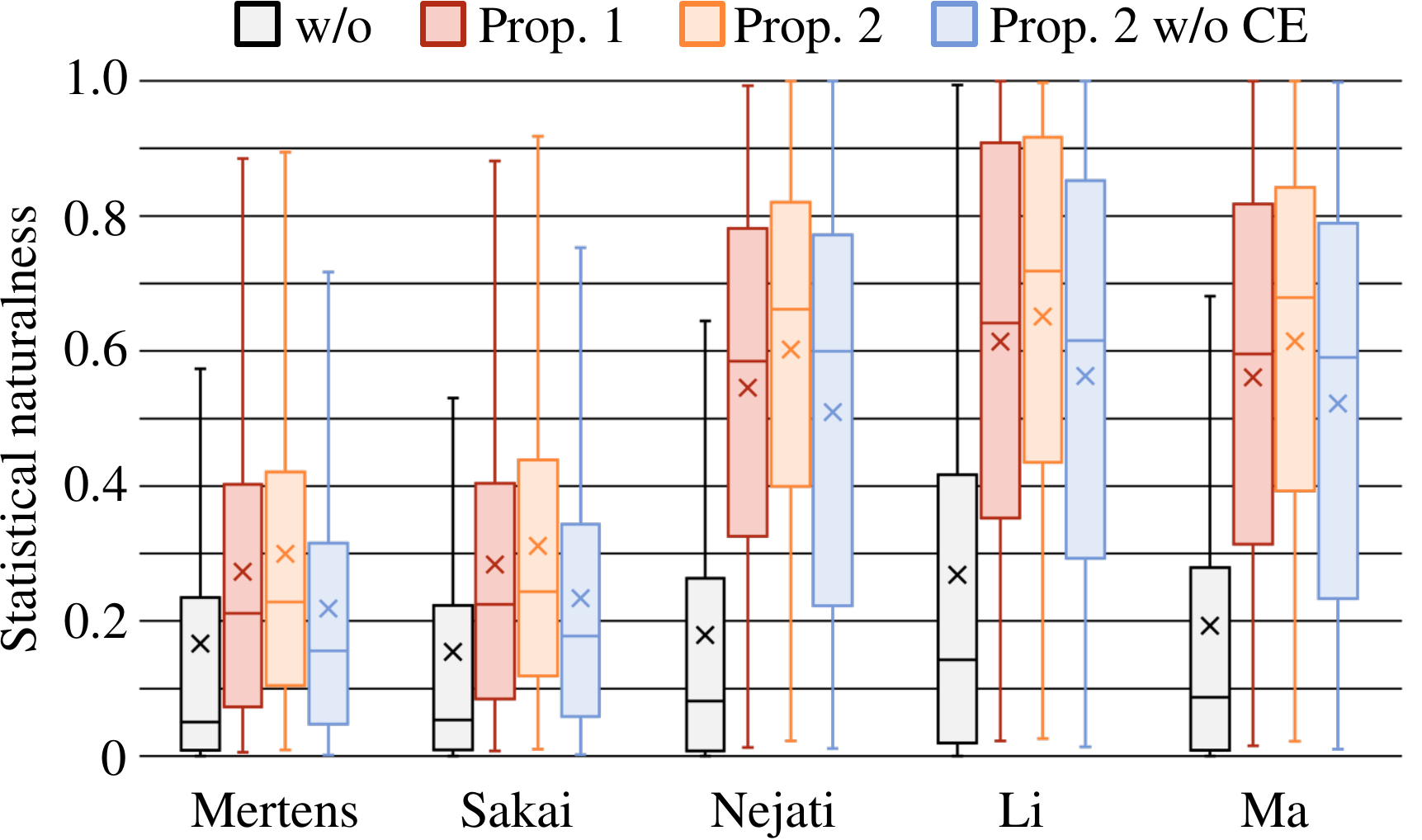}
    \caption{Statistical naturalness
      \label{fig:naturalness_random}}
  \end{subfigure}
  \begin{subfigure}[t]{0.24\hsize}
    \centering
    \includegraphics[width=\columnwidth]{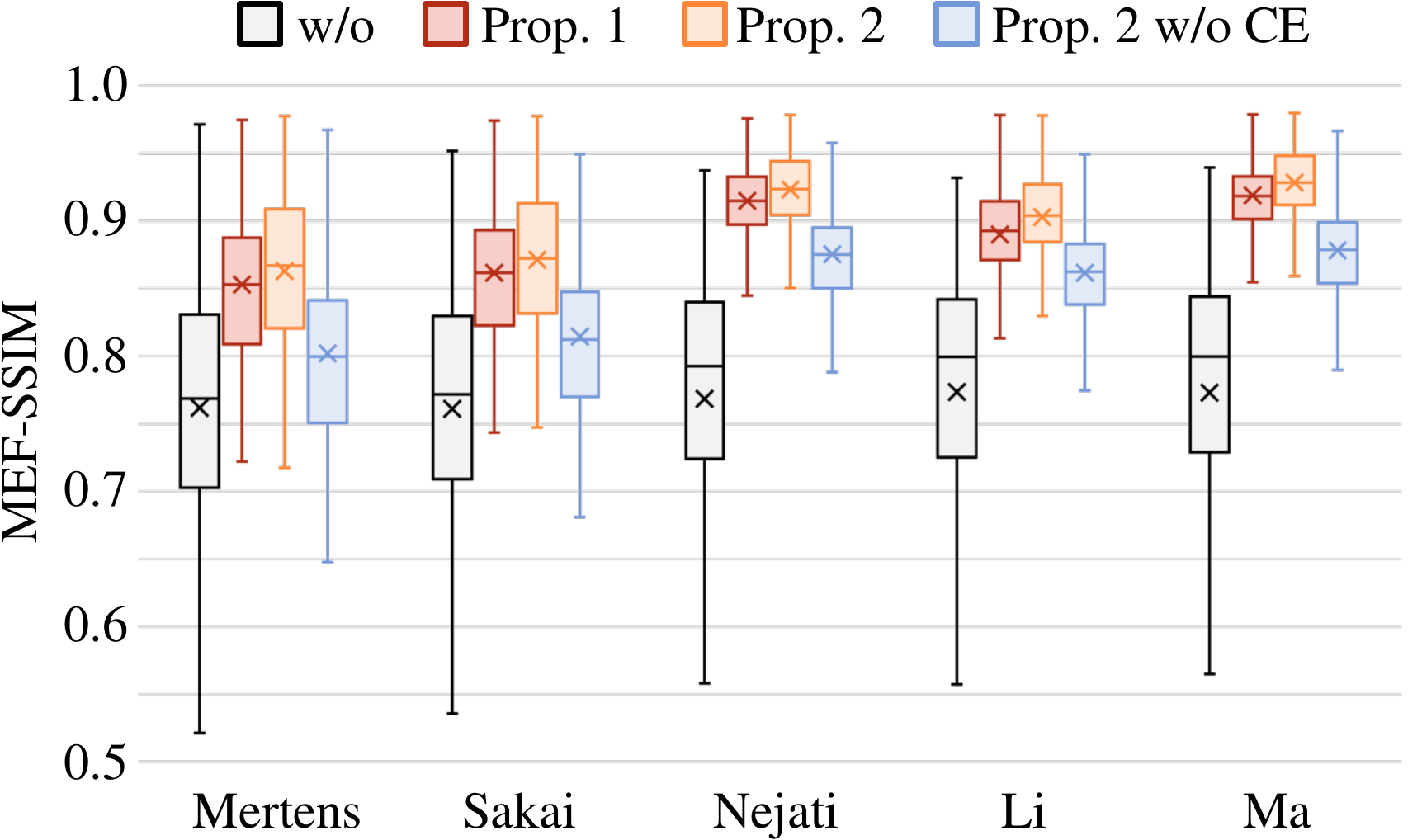}
    \caption{MEF-SSIM
      \label{fig:mefssim_random}}
  \end{subfigure}
  \caption{Quantitative evaluation for 500 sets of tone-mapped images.
    ``w/o'' means that images $\myvector{y}$ were produced without luminance adjustment.
    ``Prop. 1'' and ``Prop. 2'' indicate proposed method using Approach 1 and Approach 2,
    respectively.
    ``Prop. 2 w/o CE'' indicates proposed method using Approach 2 without
    local contrast enhancement.
    Boxes span from first to third quartile, referred to as $Q_1$ and $Q_3$,
    and whiskers show maximum and minimum values in range of
    $[Q_1 - 1.5(Q_3 - Q_1), Q_3 + 1.5(Q_3 - Q_1)]$.
    Band inside boxes indicates median, and crosses denote average value.
    \label{fig:scores_random}}
\end{figure*}

  Figure \ref{fig:scores_random} summarizes quantitative evaluation results
  for the 500 sets of unclear tone-mapped images
  in terms of discrete entropy, TMQI, statistical naturalness,
  and MEF-SSIM as box plots.
  The boxes span from the first to the third quartile, referred to as $Q_1$ and $Q_3$,
  and the whiskers show the maximum and minimum values in the range of
  $[Q_1 - 1.5(Q_3 - Q_1), Q_3 + 1.5(Q_3 - Q_1)]$.
  The band inside the boxes indicates the median, i.e., the second quartile $Q_2$,
  and the crosses denote the average value.
  Similarly to statistical naturalness,
  larger TMQI and MEF-SSIM scores $\in [0, 1]$ indicate higher quality.
  From this figure, it is shown that both
  of the approaches with the proposed method enable
  us to obtain images with higher-quality than images without any adjustment
  with any fusion method.
  In particular, Approach 2 provided the highest score
  under the use of the same fusion method
  in terms of all metrics.
  Therefore, the proposed SSLA is effective for multi-exposure image fusion
  when input multi-exposure images do not have
  a sufficient number of different exposure levels.

\begin{figure}[!t]
  \centering
  \includegraphics[width=\columnwidth]{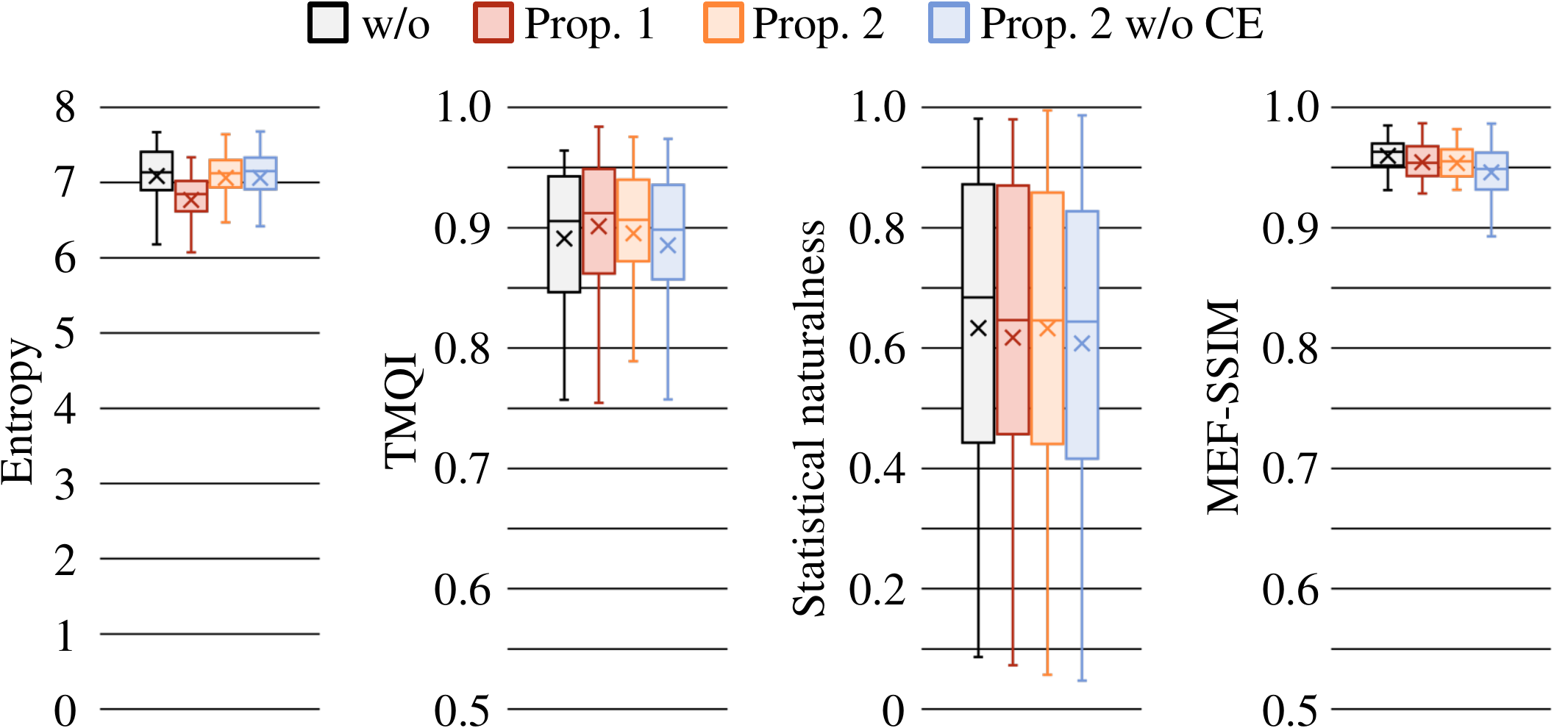}
  \caption{Quantitative evaluation for appropriate 50 sets of tone-mapped images
    (containing sufficient number of exposures) under the use of Ma's fusion method.
    ``w/o'' means that images $\myvector{y}$ were produced without luminance adjustment.
    ``Prop. 1'' and ``Prop. 2'' indicate proposed method using Approach 1 and Approach 2,
    respectively.
    ``Prop. 2 w/o CE'' indicates proposed method using Approach 2 without
    local contrast enhancement.
    Boxes span from first to third quartile, referred to as $Q_1$ and $Q_3$,
    and whiskers show maximum and minimum values in range of
    $[Q_1 - 1.5(Q_3 - Q_1), Q_3 + 1.5(Q_3 - Q_1)]$.
    Band inside boxes indicates median, and crosses denote average value.}
    \label{fig:scores_mapped}
\end{figure}

  Figure \ref{fig:scores_mapped} shows the performance of the proposed method
  when a sufficient number of multi-exposure images were given as inputs.
  Scores provided by the proposed method were
  almost the same as those of Ma's method,
  but the scores slightly less than those of Ma's method,
  Hence, it is confirmed that the proposed method does not generate a harmful effect
  even when well-exposed multi-exposure images are given.

  For these reasons,
  it was confirmed that the proposed SSLA is effective for multi-exposure image fusion.
  It is also useful for producing high quality images
  that clearly represent an entire scene.
  Comparing Approaches 1 and 2,
  Approach 2 generated clear images,
  while Approach 1 can be performed by using a closed-form.
\subsection{Computational complexity}
  To evaluate the computational complexity of the proposed method,
  we measured the executing time of the proposed one.
  Eight image sets selected from an available online database \cite{easyhdr}
  were utilized for the evaluation.
  Here, the minimum and maximum number of pixels in the image sets were
  $425,430$ and $918,400$, where their average was $533,606.3$.

  The simulation was run on a PC,
  with a 4.2 GHz processor and a main memory of 64 Gbytes (see Table \ref{tab:machine_spec}).
  In the simulation,
  \textit{timeit()} function in MATLAB was used for measuring the executing time,
  and every method was carried out on a single thread of execution.
\begin{table}[!t]
  \centering
  \caption{Machine spec used for evaluating executing time}
  \begin{tabular}{ll}\hline\hline
    Processor & Intel Core i7-7700K 4.20 GHz\\
    Memory & 64 GB\\
    OS & Ubuntu 16.04 LTS\\
    Software & MATLAB R2017b\\
    \hline
  \end{tabular}
  \label{tab:machine_spec}
\end{table}
  
\begin{figure}[!t]
  \centering
  \includegraphics[width=\columnwidth]{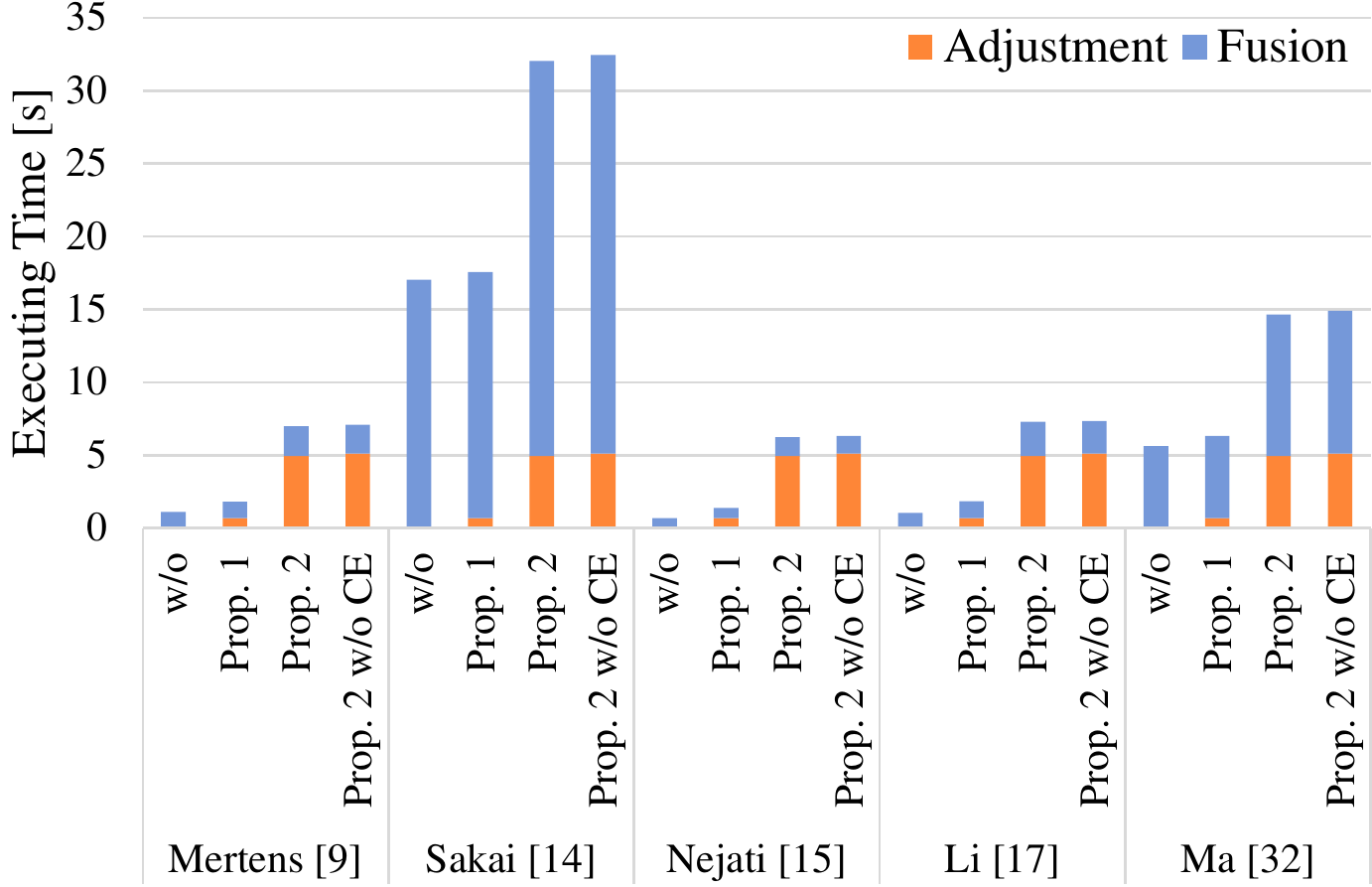}
  \caption{Average executing time for eight image sets.
    ''w/o'' means that images $\myvector{y}$ were produced without luminance adjustment.
    ``Prop. 1'' and ''Prop. 2'' indicate proposed method using Approach 1 and Approach 2,
    respectively.
    ''Prop. 2 w/o CE'' indicates proposed method using Approach 2 without
    local contrast enhancement.
    Approach 1 is faster than Approach 2.
    \label{fig:runtime}}
\end{figure}

  Figure \ref{fig:runtime} shows average executing time for the eight image sets.
  From the figure, we can see that the overhead of Approach 1 was about one second.
  In contrast, Approach 2 needs additional computational cost for adjustment
  in addition to the cost of Approach 1.
  Compared with Approach 1,
  the computational cost for fusion also increased because
  the number of adjusted multi-exposure images is generally larger than
  that of input multi-exposure images.
  Therefore, Approach 1 has a lower computational cost than Approach 2,
  while Approach 2 can generate better images than Approach 1.

\section{Conclusion}
  In this paper, we proposed a novel luminance adjustment method
  based on scene segmentation for multi-exposure image fusion.
  We first pointed out that adjusting the luminance of input images makes it possible
  to improve the quality of final fused images,
  although existing fusion methods directly fuse input multi-exposure images
  without any adjustment.
  The proposed method enables us to produce high-quality images
  even when undesirable inputs are given.
  In the method, we utilize scene segmentation
  in order to automatically adjust input multi-exposure images
  so that they become suitable for multi-exposure image fusion.
  For the scene segmentation, two approaches, 1 and 2, were proposed.
  Approach 1 separates a scene in multi-exposure images
  into a number of areas with closed-form calculation.
  In Approach 2, a scene is separated by a GMM of luminance distribution.
  In simulations, the proposed SSLA was applied to five fusion methods,
  Mertens's one, Sakai's one, Nejati's one, Li's one, and Ma's one.
  The results showed that
  the proposed SSLA is effective with any of the fusion methods
  in terms of entropy and TMQI.
  Moreover, visual comparison results showed that
  the proposed SSLA makes it possible to clearly represent
  shadow areas in images while maintaining the quality of bright areas.
  In particular, Approach 2 can generate clearer images than Approach 1.
  The proposed SSLA was also shown to be still effective for scenes
  including moving objects,
  and ghost-like artifacts were eliminated by using the proposed SSLA
  with Ma's fusion method.
  This indicates that the proposed SSLA is expected to improve
  the performance of image alignment in various image fusion problems, such as
  multi-exposure image fusion, HDR imaging, and panoramic image stitching.
  Furthermore, combining the proposed SSLA with the burst photography
  would enable us to improve the noise robustness.


%
\appendices
\section{Relationship between exposure values and pixel values \label{sec:relationship}}
\begin{figure}[!t]
    \centering
    \includegraphics[width=0.95\linewidth]{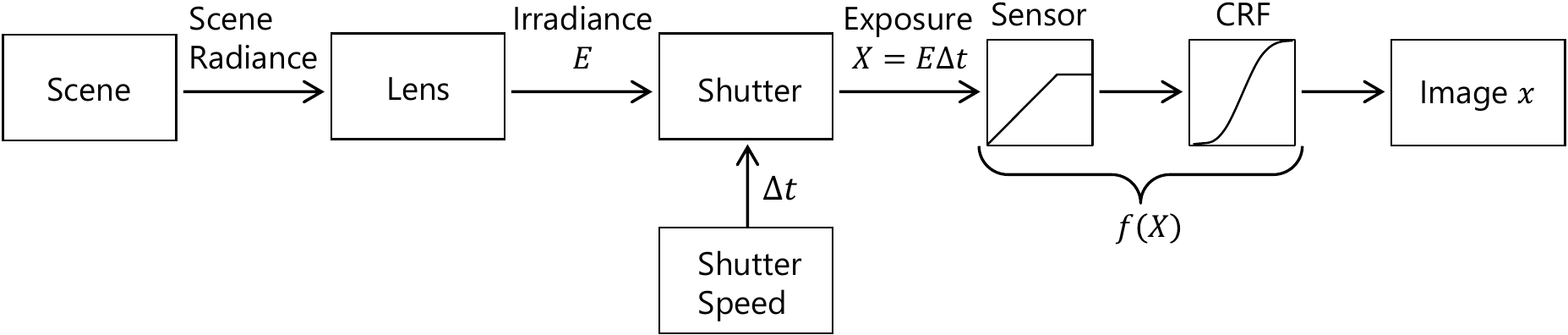}
    \caption{Imaging pipeline of digital camera \label{fig:camera}}
  \end{figure}

  Figure \ref{fig:camera} shows a typical imaging pipeline for
  a digital camera\cite{dufaux2016high}.
  Here, we focus on grayscale images for simplicity.
  The radiant power density at a sensor, i.e., irradiance $E$,
  is integrated over the time $\Delta t \in [0, \infty)$ that the shutter is open,
  producing an energy density, commonly referred to as ``exposure $X$.''
  If the scene is static during this integration,
  exposure $X$ can be written simply as the product of irradiance $E$ and
  integration time $\Delta t$ (referred to as ``shutter speed''):
  \begin{equation}
    X(\myvector{p}) = E(\myvector{p}) \Delta t.
    \label{eq:exposure}
  \end{equation}
  The photographed image $x$ is given by
  \begin{equation}
    x(\myvector{p}) = f(X(\myvector{p})),
    \label{eq:CRF}
  \end{equation}
  where $f$ is a function combining sensor saturation
  and a camera response function (CRF).
  The CRF represents the processing in each camera that makes the final image $x$
  look better.

  Camera parameters, such as shutter speed and lens aperture,
  are usually calibrated in terms of exposure value (EV) units,
  and the proper exposure for a scene is automatically decided by
  the camera.
  The exposure value is commonly controlled by changing the shutter speed
  although it can also be controlled by adjusting various camera parameters.
  Here we assume that the camera parameters except for
  the shutter speed are fixed.
  Let $v_0 = 0 \mathrm{[EV]}$ and $\Delta t_0$
  be the proper exposure value
  and shutter speed under the given conditions, respectively.
  The exposure value $v_i \mathrm{[EV]}$ of an image taken at
  shutter speed $\Delta t_i$ is given by
  \begin{equation}
    v_i = \log_2 \Delta t_i - \log_2 \Delta t_0.
    \label{eq:EV}
  \end{equation}
  From (\ref{eq:exposure}) to (\ref{eq:EV}),
  images $x_0$ and $x_i$ exposed at $0 \mathrm{[EV]}$ and $v_i \mathrm{[EV]}$,
  respectively, are written as
  \begin{align}
    x_0(\myvector{p}) &= f(E(\myvector{p})\Delta t_0) \label{eq:CRFwithExposure}\\
    x_i(\myvector{p}) &= f(E(\myvector{p})\Delta t_i)
            = f(2^{v_i} E(\myvector{p})\Delta t_0). \label{eq:CRFwithExposure2}
  \end{align}
  Assuming function $f$ is linear,
  we obtain a relationship between $x_0$ and $x_i$:
  \begin{equation}
    x_i(\myvector{p}) = 2^{v_i} x_0(\myvector{p}).
    \label{eq:relationship}
  \end{equation}
  Therefore, the exposure can be varied artificially by multiplying $x_0$ by a constant.

\section{Physical meaning of scaling RGB values \label{sec:color}}
  In the proposed SSLA,
  RGB pixel values of an adjusted multi-exposure image are given by eq. (\ref{eq:color}).
  Here, we prove that scaling RGB pixel values does not change colors of objects.
  
  Let spectral distribution of illumination light striking an object,
  spectral reflectance of the object,
  and spectral response of RGB color filters on an imaging sensor as
  $P(\lambda), R(\lambda),$ and $H_c(\lambda) (c \in {r, g, b})$,
  where $\lambda$ means wavelength of light.
  Irradiance $E_c$ at the sensor is given as
  \begin{equation}
    E_c = \int_0^\infty P(\lambda) R(\lambda) H_c(\lambda) d\lambda.
  \end{equation}
  From eqs. (\ref{eq:exposure}) and (\ref{eq:CRF}), RGB pixel value $x_c$ is written as
  \begin{align}
    x_c &= f(E_c\Delta t) \\
        &= f\left( \Delta t \int_0^\infty P(\lambda) R(\lambda) H_c(\lambda) d\lambda \right).
  \end{align}

  If the radiance power of $P(\lambda)$ is scaled with $\alpha$,
  RGB pixel value $x'_c$ is calculated by
  \begin{align}
    x'_c &= f
            \left(
              \Delta t \int_0^\infty \alpha P(\lambda) R(\lambda) H_c(\lambda) d\lambda
            \right)\\
         &= f
            \left(
              \alpha \Delta t \int_0^\infty P(\lambda) R(\lambda) H_c(\lambda) d\lambda
            \right).
  \end{align}
  Assuming function $f$ is linear, namely, $x_c$ is a linear-RGB pixel value,
  we obtain a relationship between $x_c$ and $x'_c$:
  \begin{equation}
    x'_c = \alpha x_c.
  \end{equation}
  Therefore, scaling RGB pixel value $x_c$ with $\alpha$ corresponds to
  scaling the power of illumination light with $\alpha$.
  For this reason,
  scaling RGB pixel values does not change colors of objects.

\section*{Acknowledgment}
  This work was suppoeted by JSPS KAKENHI Grant Number JP18J20326.

\ifCLASSOPTIONcaptionsoff
  \newpage
\fi




%

\begin{IEEEbiography}[{\includegraphics[width=1in,height=1.25in,clip,keepaspectratio]{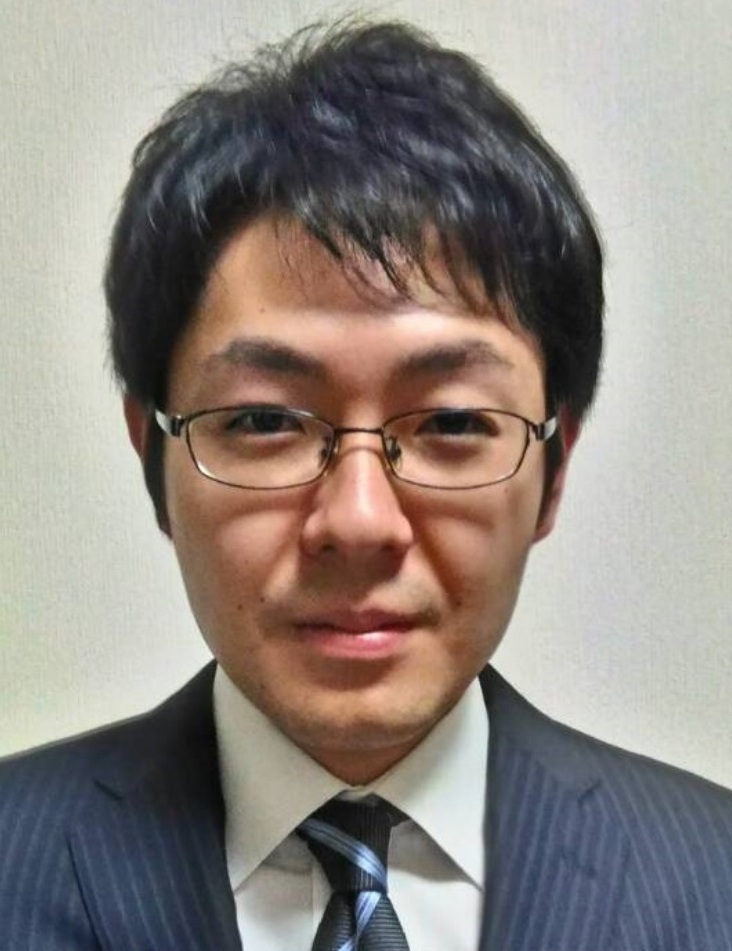}}]
{Yuma Kinoshita}
  received his B.Eng. and M.Eng. degrees from Tokyo Metropolitan University,
  Japan, in 2016 and 2018, respectively.
  From 2018, he has been a Ph.D. student
  at Tokyo Metropolitan University.
  He received IEEE ISPACS Best Paper Award in 2016,
  IEEE Signal Processing Society Japan Student Conference Paper Award in 2018,
  and IEEE Signal Processing Society Tokyo Joint Chapter Student Award in 2018,
  respectively.
  His research interests are in the area of image processing.
  He is a student member of IEEE and IEICE.
\end{IEEEbiography}

\begin{IEEEbiography}[{\includegraphics[width=1in,height=1.25in,clip,keepaspectratio]{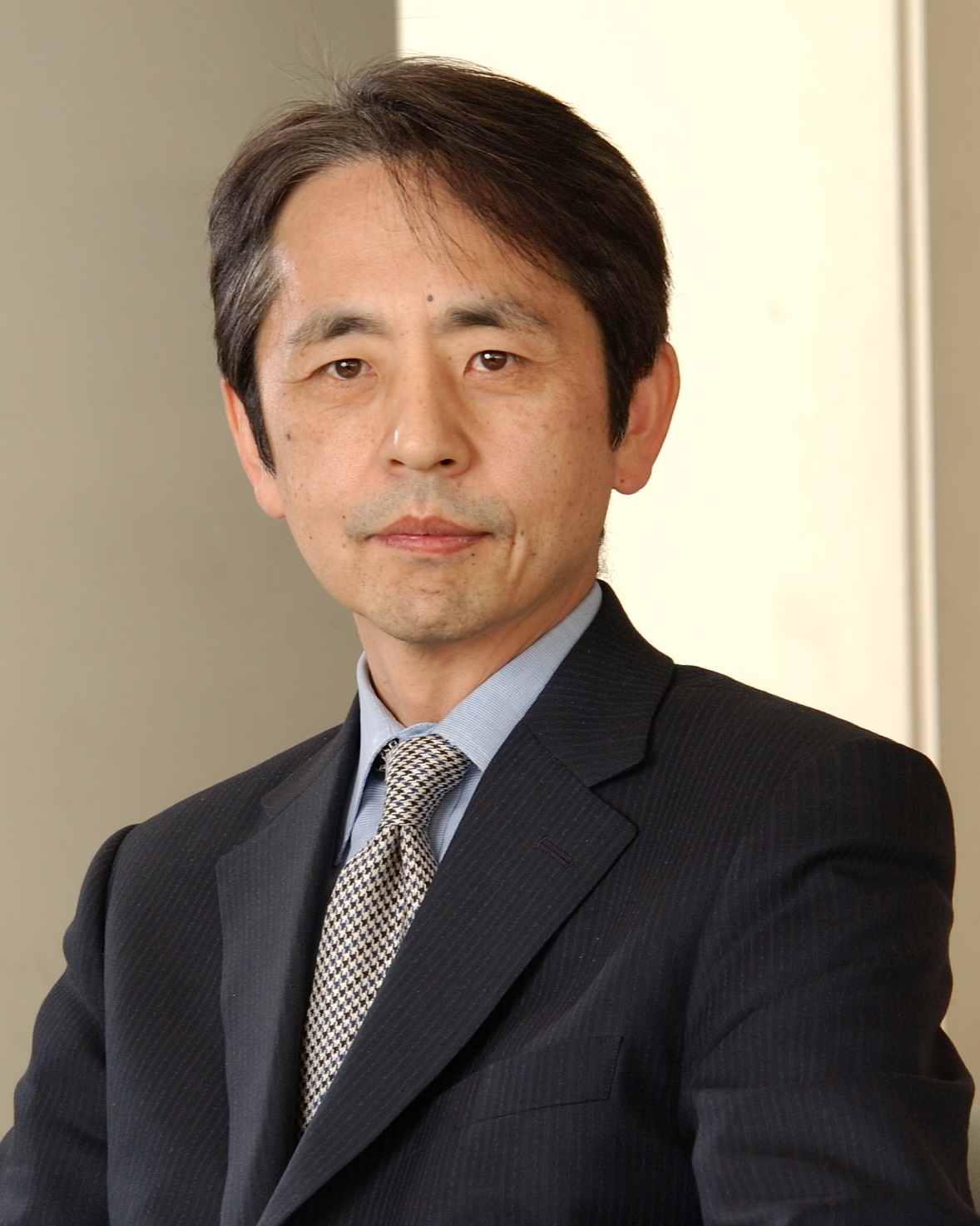}}]
{Hitoshi Kiya}
  received his B.E and M.E. degrees from Nagaoka University of Technology,
  in 1980 and 1982 respectively, and his Dr. Eng. degree
  from Tokyo Metropolitan University in 1987.
  In 1982, he joined Tokyo Metropolitan University, where he became Full Professor in 2000.
  From 1995 to 1996, he attended the University of Sydney, Australia as a Visiting Fellow.
  He is a Fellow of IEEE, IEICE and ITE.
  He currently serves as President of APSIPA,
  and he served as Inaugural Vice President (Technical Activities) of APSIPA in 2009-2013,
  and as Regional Director-at-Large for Region 10 of IEEE Signal Processing Society
  in 2016-2017.
  He was also President of IEICE Engineering Sciences Society in 2011-2012,
  and he served there as Vice President and Editor-in-Chief
  for IEICE Society Magazine and Society Publications.
  He was Editorial Board Member of eight journals, including IEEE Trans. on Signal Processing,
  Image Processing, and Information Forensics and Security,
  Chair of two technical committees and Member of nine technical committees
  including APSIPA Image, Video, and Multimedia Technical Committee (TC),
  and IEEE Information Forensics and Security TC.
  He has organized a lot of international conferences,
  in such roles as TPC Chair of IEEE ICASSP 2012 and as General Co-Chair of IEEE ISCAS 2019.
  Dr. Kiya is a recipient of numerous awards, including six best paper awards. 
\end{IEEEbiography}





\end{document}